\documentclass[journal=jacsat,manuscript=article]{achemso}
\setkeys{acs}{articletitle=true, maxauthors=5}

\makeatletter
\renewcommand*{\acs@author@fnsymbol@symbol}[1]{%
\ifcase #1 *\or
    \dagger\or
     \S\or
      \ddagger\or
          \|\or
           \#\or
            \triangle\or
            \bot\or
              \P\or
                \triangle\or
                  \nabla\or
                     @
    \fi
}

\makeatother

\usepackage[version=3]{mhchem} 
\usepackage[]{mathrsfs}
\usepackage[]{listings}
\usepackage[latin1]{inputenc}
\usepackage{amssymb}
\usepackage{epsfig}
\usepackage{xspace}
\usepackage{pstricks}
\usepackage{psfrag}
\usepackage{color}
\usepackage{siunitx}
\usepackage{subcaption}
\usepackage{booktabs}
\author{\vspace*{-0.1cm}\it Theresia~Knobloch}
\affiliation{~Institute for Microelectronics, TU Wien, Gusshausstrasse 27--29, 1040 Vienna, Austria}
\altaffiliation{Contributed equally to this work}
\author{\vspace*{-0.1cm}\it Burkay~Uzlu}
\affiliation{~AMO GmbH, Otto-Blumenthal-Strasse 25, 42074 Aachen, Germany}
\alsoaffiliation{~Chair of Electronic Devices, RWTH Aachen University, 52074, Aachen, Germany}
\altaffiliation{Contributed equally to this work}
\author{\vspace*{-0.1cm}\it Yury~Yu.~Illarionov}
\affiliation{~Institute for Microelectronics, TU Wien, Gusshausstrasse 27--29, 1040 Vienna, Austria}
\alsoaffiliation{ Ioffe Physical-Technical Institute, Polytechnicheskaya 26, 194021 St-Petersburg, Russia}
\email{illarionov@iue.tuwien.ac.at}
\author{\vspace*{-0.1cm}\it Zhenxing~Wang}
\affiliation{~AMO GmbH, Otto-Blumenthal-Strasse 25, 42074 Aachen, Germany}
\author{\vspace*{-0.1cm}\it Martin~Otto}
\affiliation{~AMO GmbH, Otto-Blumenthal-Strasse 25, 42074 Aachen, Germany}
\author{\vspace*{-0.1cm}\it Lado Filipovic}
\affiliation{~Institute for Microelectronics, TU Wien, Gusshausstrasse 27--29, 1040 Vienna, Austria}
\author{\it Michael~Waltl}
\affiliation{~Christian Doppler Laboratory for Single-Defect Spectroscopy in Semiconductor Devices at the Institute for Microelectronics, TU Wien, Vienna, Austria}
\author{\it Daniel~Neumaier}
\affiliation{~AMO GmbH, Otto-Blumenthal-Strasse 25, 42074 Aachen, Germany}
\alsoaffiliation{Chair of Smart Sensor Systems, University of Wuppertal, 42119 Wuppertal, Germany}
\author{\vspace*{-0.1cm}\it Max~C.~Lemme}
\affiliation{~AMO GmbH, Otto-Blumenthal-Strasse 25, 42074 Aachen, Germany}
\alsoaffiliation{~Chair of Electronic Devices, RWTH Aachen University, 52074, Aachen, Germany}
\author{\vspace*{-0.1cm}\it Tibor~Grasser}
\affiliation{~Institute for Microelectronics, TU Wien, Gusshausstrasse 27--29, 1040 Vienna, Austria}
\email{grasser@iue.tuwien.ac.at}

\title[Fermi-Level Tuning for Stable 2D Devices]
{Optimizing the Stability of FETs Based on Two-Dimensional Materials \\ by Fermi Level Tuning}

\begin{document}


\vspace*{-0.8cm}
\begin{abstract}
\vspace*{-0.1cm}
\small

Despite the enormous progress achieved during the past decade, nanoelectronic devices based on two-dimensional (2D) semiconductors still suffer from a limited electrical stability. This limited stability has been shown to result from the interaction of charge carriers originating from the 2D semiconductors with defects in the surrounding insulating materials. The resulting dynamically trapped charges are particularly relevant in field effect transistors (FETs) and can lead to a large hysteresis, which endangers stable circuit operation. Based on the notion that charge trapping is highly sensitive to the energetic alignment of the channel Fermi-level with the defect band in the insulator, we propose to optimize device stability by deliberately tuning the channel Fermi-level. Our approach aims to minimize the amount of electrically active border traps without modifying the total number of traps in the insulator. We demonstrate the applicability of this idea by using two differently doped graphene layers in otherwise identical FETs with Al$_2$O$_3$ as a gate oxide mounted on a flexible substrate. Our results clearly show that by increasing the distance of the Fermi-level to the defect band, the hysteresis is significantly reduced. Furthermore, since long-term reliability is also very sensitive to trapped charges, a corresponding improvement in reliability is both expected theoretically and demonstrated experimentally. Our study paves the way for the construction of more stable and reliable 2D FETs in which the channel material is carefully chosen and tuned to maximize the energetic distance between charge carriers in the channel and the defect bands in the insulator employed.

\end{abstract}
\textbf{Keywords:} Fermi-level tuning, field-effect transistor, oxide defects, defect bands, 2d materials, graphene, hysteresis, bias-temperature instability, reliability


Two-dimensional(2D) semiconductors hold the promise of revolutionizing nanoelectronics. Their inherent atomic layer thinness makes them a plausible candidate for ultimately scaled field-effect transistors(FETs) at the end of the roadmap of silicon technology~\cite{Akinwande2019}. In contrast to silicon, 2D semiconductors retain sizable mobilities at thicknesses below \SI{1}{nm}~\cite{English2016}, a thickness which would also suppress short-channel effects in FETs, thereby allowing for channel lengths $L<\SI{5}{nm}$~\cite{Fiori2014}. In addition, the flexible integration of 2D materials in van der Waals heterostructures~\cite{Liu2019d} opens up new design options for highly energy efficient transistors which overcome the limitations of thermal charge carrier injection~\cite{Iannaccone2018}. Beyond advancing modern nanoelectronics, 2D materials can be used for many other applications, from photonics and optoelectronics~\cite{Mak2016} over neuromorphic computing~\cite{Sangwan2020} to nano-electro-mechanical systems (NEMS)~\cite{Lemme2020}, radio-frequency devices~\cite{Schwierz2015}, Hall sensors~\cite{Wang2016f} and various gas and biological sensors~\cite{MortazaviZanjani2017}.

Overall, theoretical prospects and available prototypes indicate a bright future for 2D material based devices. Nevertheless, all application scenarios depend on the requirement that devices need to show stable operation throughout their lifetime, as defined by the stability of the threshold voltage $V_\mathrm{TH}$. In graphene FETs (GFETs) the threshold voltage corresponds to the charge neutrality or Dirac voltage ($V_\mathrm{Dirac}$), as the gate voltage where the current is at its minimum~\cite{Martin2008}. When using FETs as switches in digital logic, the circuitry inherently relies on stable $V_\mathrm{TH}$ of all FETs it comprises. Also applications in radio-frequency electronics rely on a stable operating point and in gas sensors for example the shift of $V_\mathrm{TH}/ V_\mathrm{Dirac}$ can serve as measurement signal~\cite{Fu2017a}, thus unrelated drifts of these properties lead to measurement errors. In consequence, it is essential that FETs show minimal instabilities of the threshold voltage regardless of previous biasing, switching frequencies or temperatures. 

However, stability studies on 2D FETs are scarce and typically show that their stability is at least two orders of magnitude worse~\cite{Yang2014c, Illarionov2017c} compared to silicon based FETs~\cite{Stathis2006, Grasser2011b}. 
Typical measurements of FET stability are the evaluation of the hysteresis in the transfer characteristics~\cite{Late2012} and of the stability of the threshold voltage under prolonged periods of applied elevated gate biases and temperatures, the bias temperature instability~(BTI)~\cite{Stathis2006}. As a root cause for these phenomena early on charge trapping inside the gate oxide has been identified~\cite{Thomas1964}, where, facilitated by elevated gate biases and temperatures, charges are transferred between the channel and the gate oxide in a phonon mediated transition with charging time constants spanning a wide range from ns up to years~\cite{Grasser2012}.
In optimized and stable silicon FETs these border traps in the gate oxide close to the channel determine the long-term stability and reliability~\cite{Fleetwood1993, Grasser2011b}, but in 2D material based FETs border traps are responsible for limited device stability on shorter time scales~\cite{Illarionov2016f}.
In a first approximation, the energy levels of the defects follow a normal distribution around the average defect levels, forming defect bands~\cite{Kaczer2018}, as in amorphous gate oxides the local surroundings of every defect differ leading to a variation in the defects' trap levels~\cite{Goes2018}.
As a consequence, the overall density of border traps and the widths of the corresponding defect bands could be considerably reduced when using crystalline insulators, such as hexagonal boron nitride~(hBN) or calcium fluoride~(CaF\textsubscript{2})~\cite{Illarionov2020}. However, these insulators are difficult to synthesize and come with numerous technological challenges. For example, at the current state of the art crystalline hBN can only be grown at temperatures above \SI{1200}{\degree C}~\cite{Shi2020} and CaF\textsubscript{2} requires a crystalline silicon (111) substrate for growth, allowing only back-gated configurations~\cite{Illarionov2019a}. In addition, hBN is unsuitable for use as a scaled gate insulator because of its small dielectric constant~\cite{Knobloch2021}.
Therefore, it would be an important breakthrough if stable FETs based on 2D semiconductors could be built based on common amorphous gate oxides such as SiO\textsubscript{2}, Al\textsubscript{2}O\textsubscript{3} or HfO\textsubscript{2}. 

Here, we address the need of stable 2D FETs by suggesting a novel engineering approach. We aim to build stable 2D FETs by carefully selecting 2D materials and tuning their Fermi level~($E_\mathrm{F}$) such that $E_\mathrm{F}$ does not come near to any defect band in the amorphous gate oxide during device operation. This can be realized by careful selection of the 2D material and the amorphous gate oxide and by doping the layer to move $E_\mathrm{F}$ to the desired location. Our approach thus constitutes a stability-based design which targets to form a metal-oxide-semiconductor~(MOS) system with a minimal amount of \textit{electrically active} border traps without actually modifying the \textit{total number} of traps in the insulator. We demonstrate that in this way, both electrical stability and reliability of 2D material based FETs can be improved. We apply our design method to GFETs with an aluminum oxide (Al\textsubscript{2}O\textsubscript{3}) layer as gate oxide, where we tune $E_\mathrm{F}$ in one batch of devices by p-doping the graphene layer, thereby validating this approach. The proposed stability-based design could act as a game changer which might allow to fabricate stable 2D material based FETs, neuromorphic memory elements and sensors in the future.

\section{Fermi Level Tuning for Stable 2D FETs}

\begin{figure}[!htpb]
\begin{subfigure}[c]{.8\textwidth}
\caption{\footnotesize Graphene FET}
\label{fig:stab_graph}
\includegraphics[width=\textwidth]{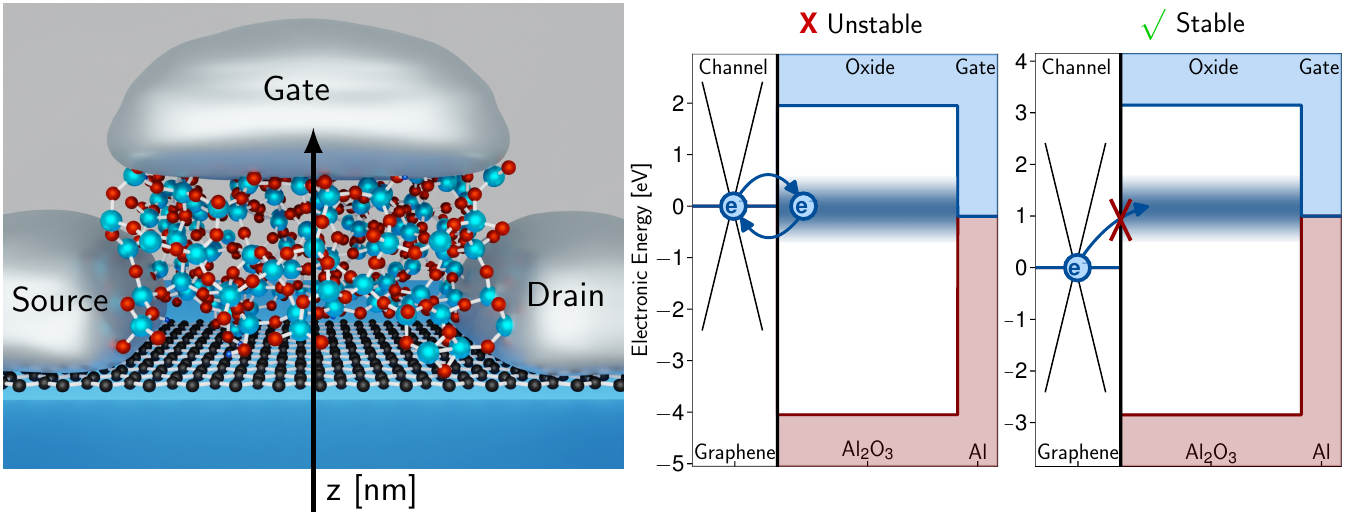}
\end{subfigure}
\begin{subfigure}[c]{.8\textwidth}
\caption{\footnotesize 2D Semiconductor FET}
\label{fig:stab_WS2}
\includegraphics[width=\textwidth]{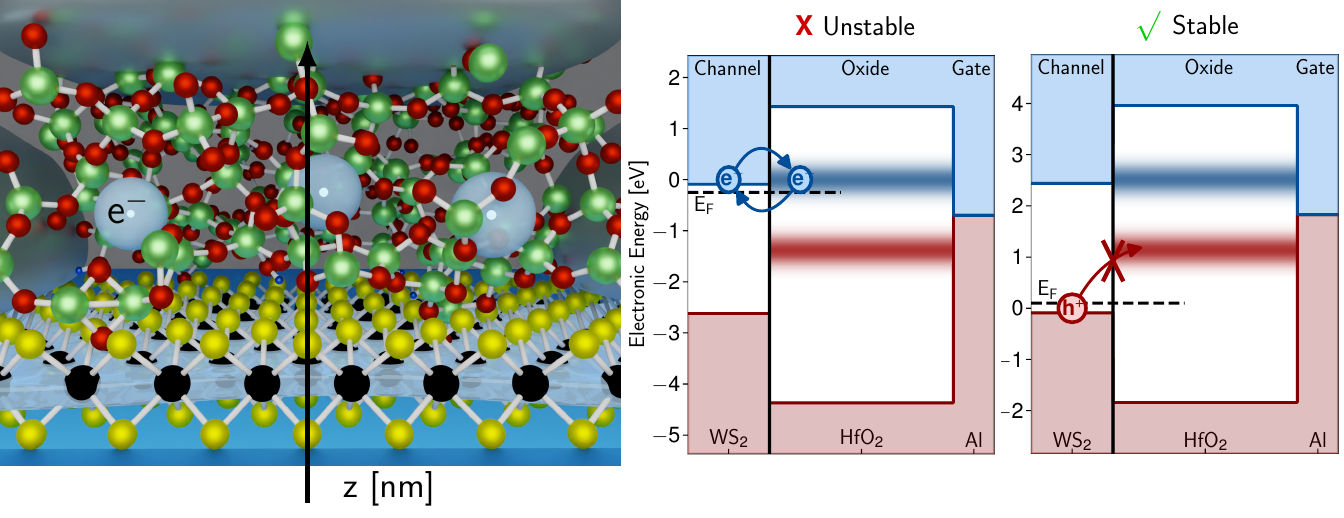}
\end{subfigure}
\begin{subfigure}[c]{.405\textwidth}
\caption{}
\label{fig:tune_graph}
\includegraphics[height=.185\textheight]{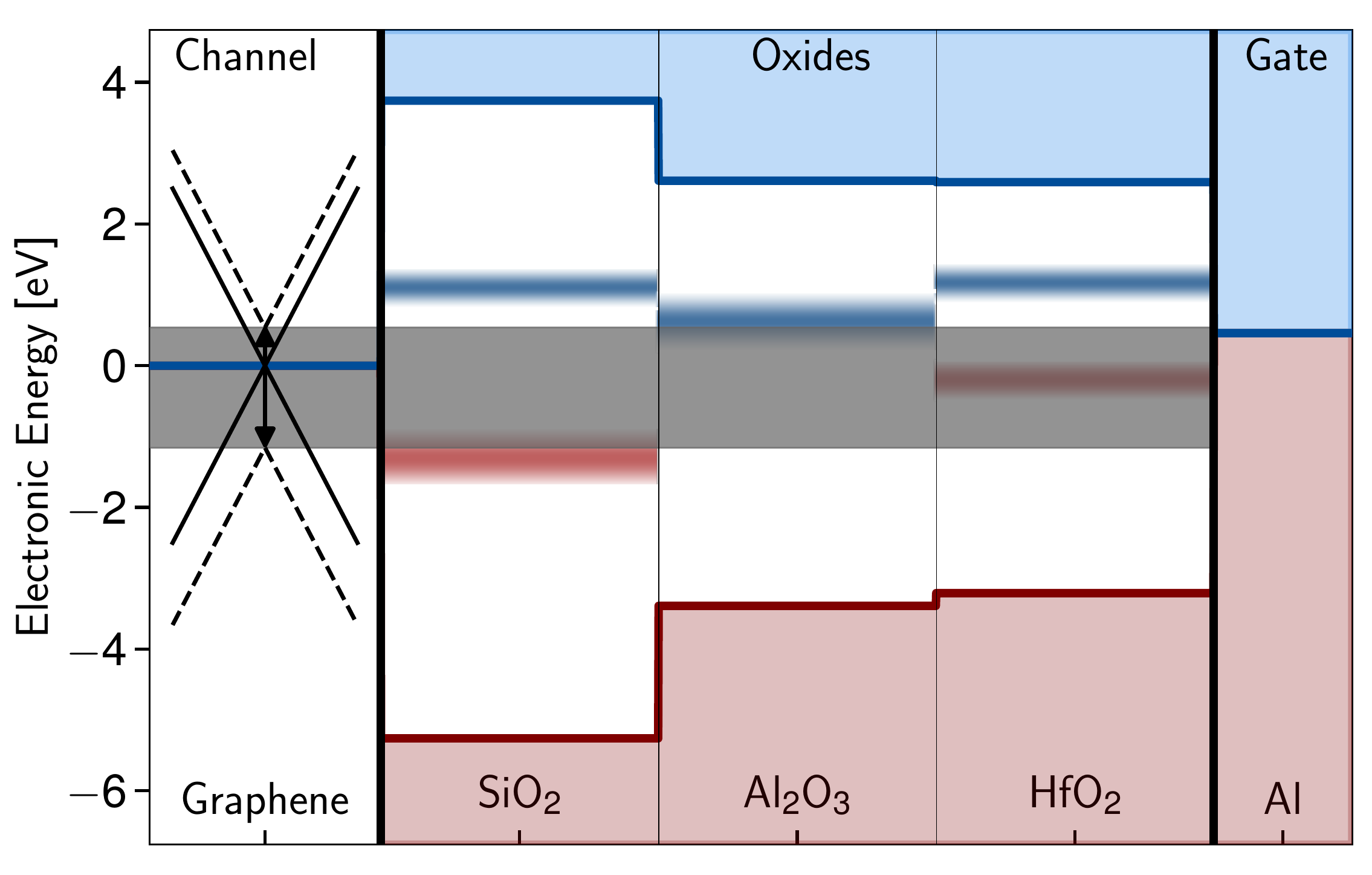}
\end{subfigure}
\begin{subfigure}[c]{.505\textwidth}
\caption{}
\label{fig:tune_WS2}
\includegraphics[height=.187\textheight]{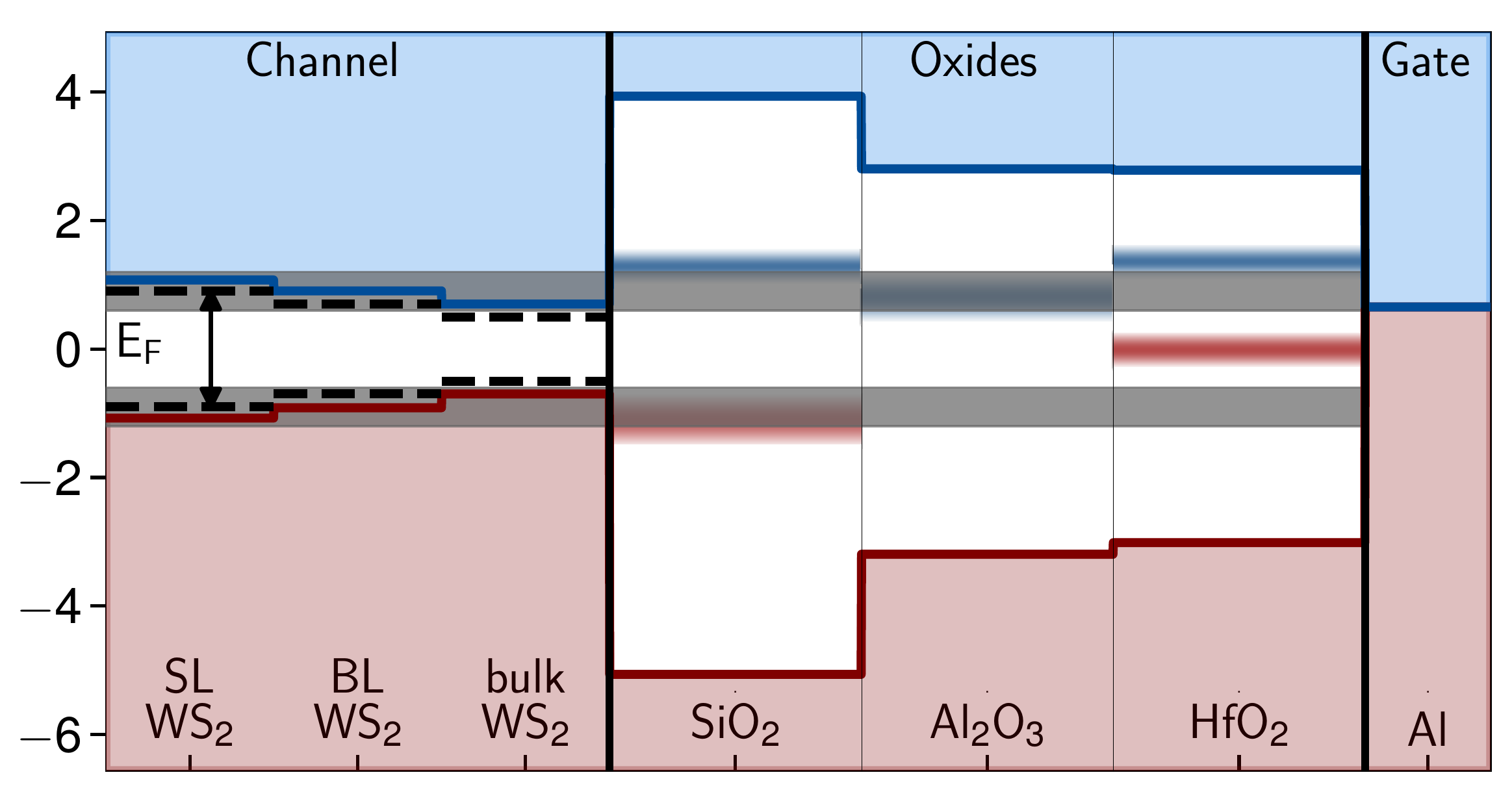}
\end{subfigure}

\caption{\footnotesize 
(a) The schematic image to the left shows a top gated GFET with an Al\textsubscript{2}O\textsubscript{3} gate oxide. For a cut through the GFET along the indicated arrow, the energetic alignment of the Fermi level to the defect band in the aluminum gate oxide is shown. In the band diagram to the left the device is electrically unstable with respect to variations in the threshold voltage as the Fermi level is aligned within the defect band. In the band diagram to the right the Fermi level has been shifted downwards rendering the device more stable.
(b) To the left the charge transfer of electrons flowing through the WS\textsubscript{2} channel to traps in the HfO\textsubscript{2} gate oxide is illustrated schematically. This situation is depicted in the left band diagram where the Fermi level is aligned close to the conduction band edge rendering the device unstable. If the Fermi level is instead aligned close to the valence band edge, the FET is stable.
(c) In this band diagram the possible range of the graphene Fermi level which is attainable via doping is shown as a grey shaded region. The Fermi level can be continuously tuned within this region.
(d) The injection of electrons and holes from the band edges of WS\textsubscript{2} is shown. 
In a semiconductor, the number of layers modifies the band gap and doping determines whether electrons or holes will be the majority carriers and thus govern device stability.
}
\end{figure}

Our stability-based design approach is based on the analysis and the design of the band diagram of the MOS system, see for example a top-gated GFET in Figure~ \ref{fig:stab_graph} which forms a MOS system out of aluminum (metal),  Al\textsubscript{2}O\textsubscript{3} (oxide) and graphene (semiconductor). By cutting through the MOS stack the corresponding band diagram is obtained, as indicated by the arrow in Figure \ref{fig:stab_graph} to the left. Every material is characterized in this view by its electron affinity, thus the energetic distance of the conduction band edge to the vacuum level and its band gap. In the case of metals and semi-metals, the work function, the energetic distance of $E_\mathrm{F}$ to the vacuum level determines the energetic location of charge carriers. At the core of our design approach lies the knowledge about the energetic position of the oxide's defect bands and their alignment to $E_\mathrm{F}$.

This energetic position of defect bands in amorphous oxides is an intrinsic material property~\cite{Blochl2000,Rzepa2018,Shluger2020}, as defect bands are related to certain defective atomic configurations inside the amorphous material which result in slightly varying trap levels depending on the local surroundings of the defects. In effect, the superposition of the trap levels of many atomic defects forms the defect band, characterized by the average energetic trap level $\overline{E_\mathrm{T}}$ and the standard deviation of the trap level distribution $\sigma_\mathrm{E_T}$. In order to experimentally determine the energetic location of defect bands, the oxide defect states can be probed by electrical measurements which analyze conductance variations in MOS systems~\cite{Degraeve2008, Nagumo2010} or by electron paramagnetic resonance measurements which detect the magnetic moment of unpaired electrons~\cite{Weeks1994}. Also, defect bands can be located with ab-initio calculations where possible defect states and their prevalence are analyzed, thereby identifying electrically active defect configurations like oxygen vacancies~\cite{MunozRamo2007} or hydrogen-related defects~\cite{Grasser2014}. Currently, the energetic locations of oxide defect bands are known for SiO\textsubscript{2}~\cite{Nagumo2010}, HfO\textsubscript{2}~\cite{MunozRamo2007, Rzepa2018} and Al\textsubscript{2}O\textsubscript{3}~\cite{Degraeve2008}. 


Based on the band alignment of the graphene work function to the defect bands in Al\textsubscript{2}O\textsubscript{3}, we can predict the electrical stability of the threshold voltage in these FETs. In the band diagram to the left of Figure \ref{fig:stab_graph}, the work function of graphene amounts to \SI{3.9}{eV}, thus graphene's $E_\mathrm{F}$ is in the middle of the Al\textsubscript{2}O\textsubscript{3} defect band. This value of $E_\mathrm{F}$ corresponds to n-doped graphene, for example using self-assembled monolayers with amine functional groups as a substrate~\cite{Park2011a}. Due to the alignment of the graphene Fermi level within the defect band, charge traps in the oxide will frequently capture and emit charges. As the applied gate voltage modifies the charging probabilities of the defects according to the electric field~\cite{Grasser2012},  $V_\mathrm{TH}$ depends on previous biasing and a pronounced hysteresis will be visible as well as considerable $V_\mathrm{TH}$ drifts during prolonged periods of applied gate biases. 

However, the FET stability can be tuned by moving $E_\mathrm{F}$ down via p-doping the graphene layer, as depicted in the band diagram to the right of Figure \ref{fig:stab_graph}. Here, $E_\mathrm{F}$ of graphene amounts to \SI{5.1}{eV}, as achieved through p-doping for example by depositing gold nanoparticles on the graphene surface~\cite{Shi2010a}. As graphene's Fermi level is located below the Al\textsubscript{2}O\textsubscript{3} defect band, charge transfer is unlikely and rare. Therefore, the oxide defects are electrically inactive, resulting in stable $V_\mathrm{TH}$ throughout device operation, independent of applied biases. In graphene, doping with different adsorbates and substrates yields a quasi-continuous variation of the Fermi level within \SI{3.4}{eV} and \SI{5.1}{eV}\cite{Kwon2012a, Wittmann2020} which can be used to tune the Fermi level during device design to minimize the impact of oxide defect bands.

When designing FETs based on 2D semiconductors, the stability-based design process needs to be adapted, see Figure \ref{fig:stab_WS2} for a schematic drawing of a WS\textsubscript{2} FET with a HfO\textsubscript{2} top gate oxide.
If the Fermi level is aligned close to the conduction band, electrons within WS\textsubscript{2} are the majority charge carriers dominating the current flow in Schottky barrier FETs~\cite{Appenzeller2016}. As the conduction band edge of WS\textsubscript{2} is aligned within the electron trapping defect band of HfO\textsubscript{2}, charge transfer to oxide defects is frequent. If WS\textsubscript{2} were p-doped instead of n-doped, holes at the valence band edge would be the majority, see the band diagram to the right of Figure \ref{fig:stab_WS2}. As the valence band edge of WS\textsubscript{2} is located below the hole trapping band in HfO\textsubscript{2}, a charging of oxide defects is very improbable. Therefore, for p-doped WS\textsubscript{2} in combination with a HfO\textsubscript{2} gate oxide there are no electrically active oxide traps, leading to a stable $V_\mathrm{TH}$ during device operation. It should be noted that for 2D semiconductors the charges are always injected from the conduction or valence band edge respectively. Thus, when designing a stable n-type or p-type MOSFET suitable combination of 2D semiconductor to oxide needs to be chosen.  

Possibilities for tuning the stability in the context of stability-aware device design are illustrated in Figures \ref{fig:tune_graph} and \ref{fig:tune_WS2}. By doping the graphene layer, graphene's $E_\mathrm{F}$ can be tuned within the whole grey shaded area in Figure \ref{fig:tune_graph}. Thus, the design freedom for stability based device design is large in graphene based FETs, and the role of SiO\textsubscript{2} defect bands can be reduced with an $E_\mathrm{F}$ alignment in the middle of the two defect bands and the impacts of the Al\textsubscript{2}O\textsubscript{3} defect band can be minimized for p-doped graphene layers. For 2D semiconductors like WS\textsubscript{2}, the design freedom for stability aware design is smaller. In Figure \ref{fig:tune_WS2} it is shown that either the conduction or the valence band edge can be chosen via doping. However, n-type WS\textsubscript{2} will presumably be electrically unstable for these three amorphous oxides and electrically stable p-type FETs could be designed using Al\textsubscript{2}O\textsubscript{3} or HfO\textsubscript{2}. 

\begin{figure}[!htb]
\centering
 \begin{subfigure}[c]{0.30\textwidth}
\caption{}
\label{fig:tune_hyst}
\centering
\includegraphics[width=\textwidth]{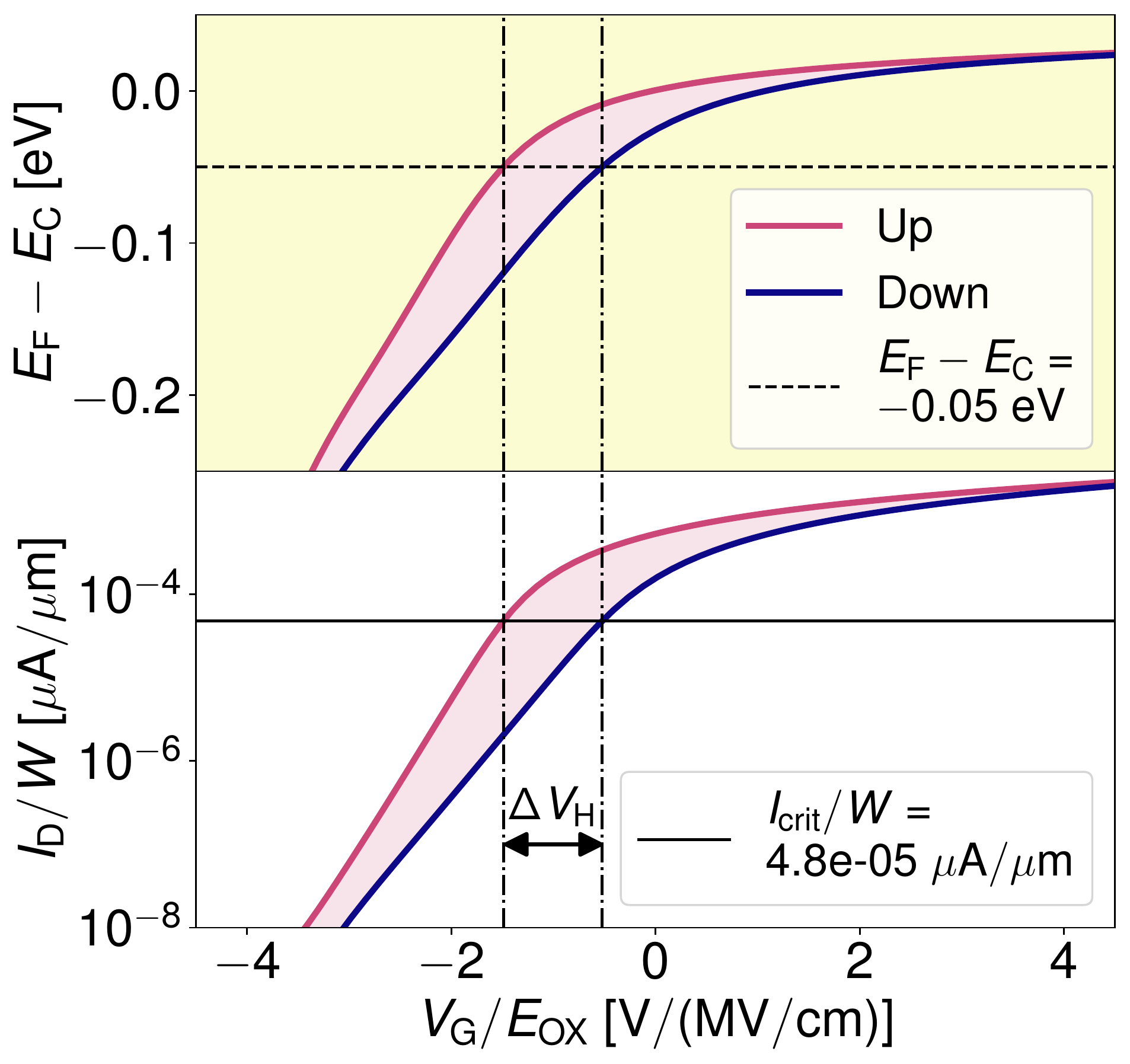}
\end{subfigure}
\begin{subfigure}[c]{0.30\textwidth}
\caption{}
\label{fig:tune_fact}
\centering
\includegraphics[width=\textwidth]{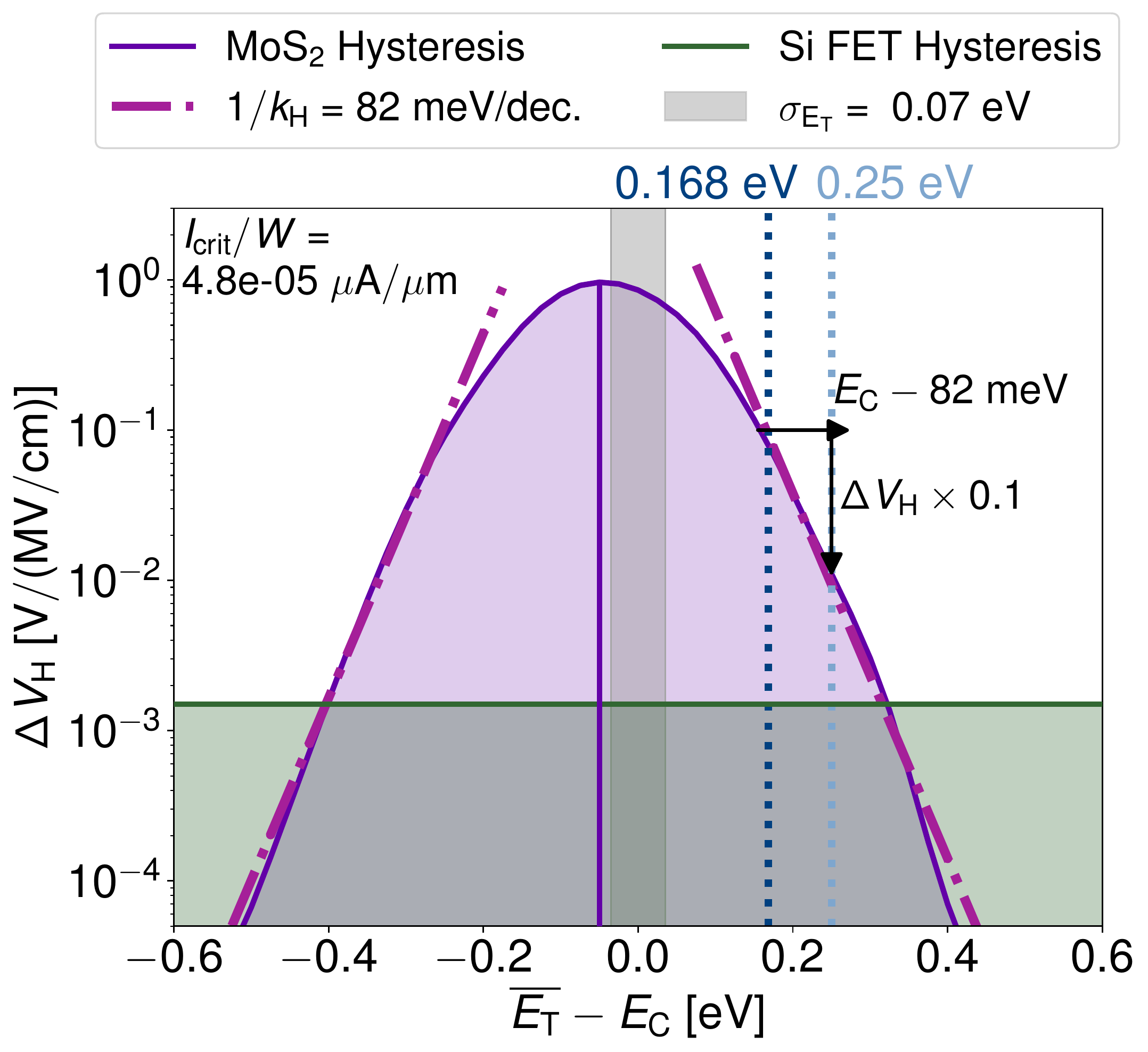}
\end{subfigure}
\begin{subfigure}[c]{0.38\textwidth}
    \caption{}
    \label{fig:tune_band}
    \begin{minipage}{0.5\textwidth}
    \includegraphics[width=\textwidth]{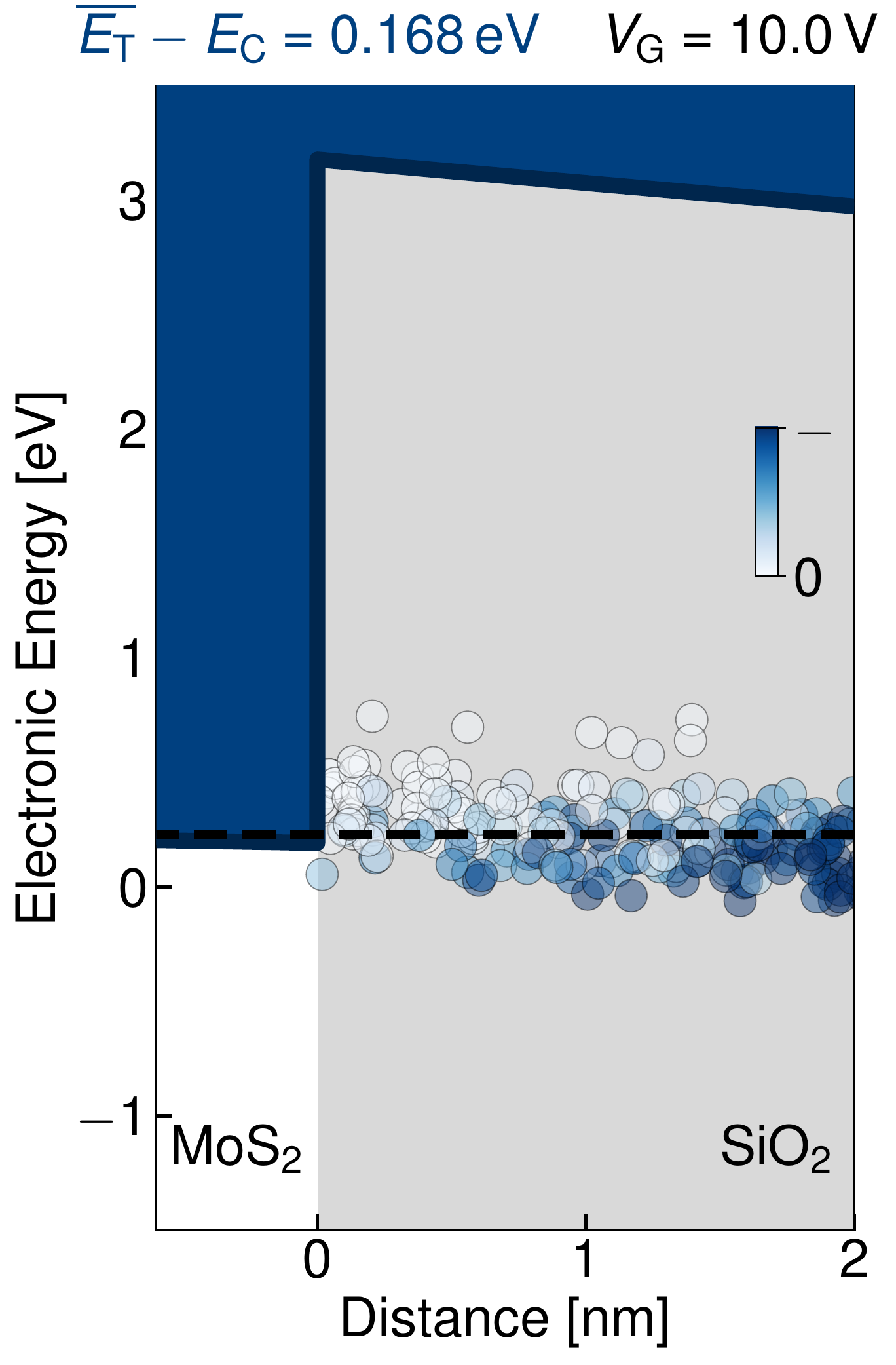}
    \end{minipage}
    \hspace*{-0.1cm}
    \begin{minipage}{0.47\textwidth}
    \includegraphics[width=\textwidth]{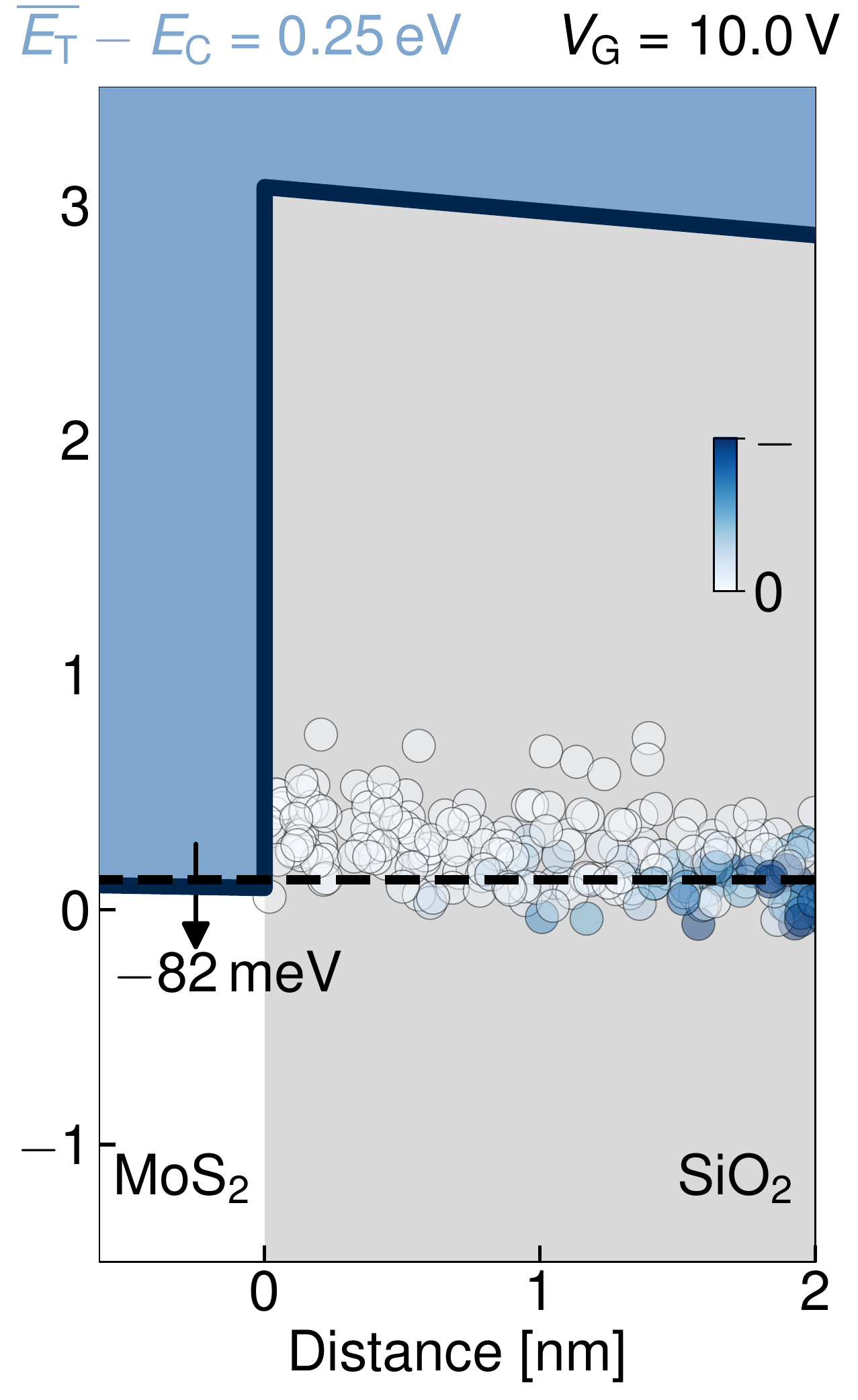}
    \end{minipage}
\end{subfigure}
\caption{
\footnotesize \textbf{(a)} At the top the calculated distance of the MoS\textsubscript{2} Fermi level to its conduction band edge is shown. The hysteresis width $\Delta V_\mathrm{H}$ is extracted at the threshold voltage, defined as $E_\mathrm{F}$ being located at \SI{50}{meV} below the conduction band edge. From simulated transfer characteristics of the MoS\textsubscript{2} FETs based on SiO\textsubscript{2}, shown below, the constant current criterion of $I_\mathrm{crit}=\SI{4.8e-5}{\micro A/\micro m}$ was used to evaluate $\Delta V_\mathrm{H}$.
\textbf{(b)} The hysteresis width $\Delta V_\mathrm{H}$ is shown on a logarithmic scale as a function of the distance of the oxide trap level $\overline{E_\mathrm{T}}$ to the MoS\textsubscript{2} conduction band edge $E_\mathrm{C}$. If $E_\mathrm{CB}$ is moved \SI{82}{meV} down, away from the trap band, the hysteresis width will improve by one order of magnitude.
\textbf{(c)} At two different locations of $E_\mathrm{C}$, namely at $\overline{E_\mathrm{T}}-E_\mathrm{C}=\SI{0.168}{eV}$ in dark blue and \SI{0.25}{eV} in light blue corresponding to the colors of the dotted lines in (b), the band diagrams of the MoS\textsubscript{2}/SiO\textsubscript{2} system are shown, demonstrating how fewer oxide traps change their charge state if the conduction band edge is shifted down, leading to a hysteresis width reduced by one order of magnitude, see (b). 
}
\label{fig:tuning}
\end{figure}

It should be noted that small energy shifts of conduction or valence band edges can be sufficient to considerably improve device stability. In Figure \ref{fig:tuning}, we used the previously developed drift-diffusion based TCAD methodology~\cite{Knobloch2018} to give an estimate for the improvement of the hysteresis width in FETs based on 2D semiconductors due to shifts of the conduction band edge $E_\mathrm{CB}$, for a detailed description of the simulation methodology see the Supporting Information (SI), Figure S1. 
We evaluated the hysteresis width at $V_\mathrm{TH}$, defined here as the voltage where the Fermi level is located only \SI{-0.05}{eV} below the conduction band edge, see Figure \ref{fig:tune_hyst} in a model system of monolayer MoS\textsubscript{2} with SiO\textsubscript{2} serving as a gate oxide. Based on the criterion for $E_\mathrm{F}-E_\mathrm{CB}$ a constant current criterion was defined and the hysteresis width was evaluated as a function of varying distance of the trap level $\overline{E_\mathrm{T}}$ to $E_\mathrm{CB}$. 
For an oxide defect band width of $\sigma_\mathrm{E_\mathrm{T}}$ of \SI{0.07}{eV} the hysteresis width can be reduced by one order of magnitude if the conduction band edge is shifted \SI{82}{meV} downwards, as illustrated in the band diagrams in Figure~\ref{fig:tune_band}. 
These small shifts of the conduction or valence band edges can be achieved by transitioning from monolayers to bulk material as shown in \ref{fig:tune_WS2}. For example in WS\textsubscript{2}, conduction and valence band edges shift by approximately \SI{160}{meV} when using bilayers instead of monolayers or by  about \SI{370}{meV} when using bulk WS\textsubscript{2} instead of monolayers, as the band gap is gradually reduced~\cite{Jo2014}. Thus, we would expect that n-type WS\textsubscript{2} FETs with an HfO\textsubscript{2} gate oxide are most stable when using bulk WS\textsubscript{2} as a channel compared to thinner WS\textsubscript{2} layers.
Independently, graphene with its continuous tunability of $E_\mathrm{F}$ over an interval of nearly \SI{2}{eV} provides the largest design freedom and because of the possibility to tune the Fermi level in graphene by a few \SI{100}{meV} through moderate doping we chose a graphene/Al\textsubscript{2}O\textsubscript{3} model system to experimentally verify our stability-based design approach.

\section{Graphene Fermi Level and Al\textsubscript{2}O\textsubscript{3} Defect Bands}

\begin{figure}[!htb]
\centering
    \hspace*{0.20cm}
    \begin{subfigure}[c]{0.61\textwidth}
        \caption{}
        \centering
        \vspace*{0.85cm}
        \includegraphics[width=0.95\textwidth]{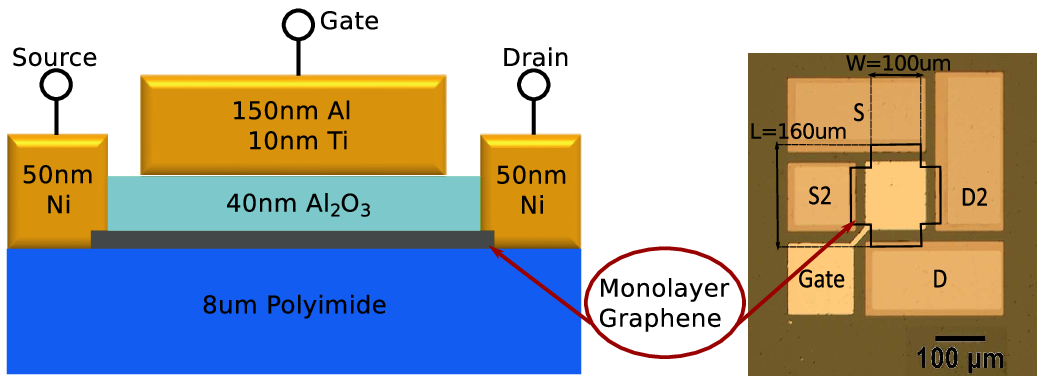}
        \vspace*{0.6cm}
        \vspace*{0.15cm}
        \label{fig:schematic}
    \end{subfigure}
    \begin{subfigure}[c]{0.305\textwidth}
        \caption{}
        \centering
        \includegraphics[width=\textwidth]{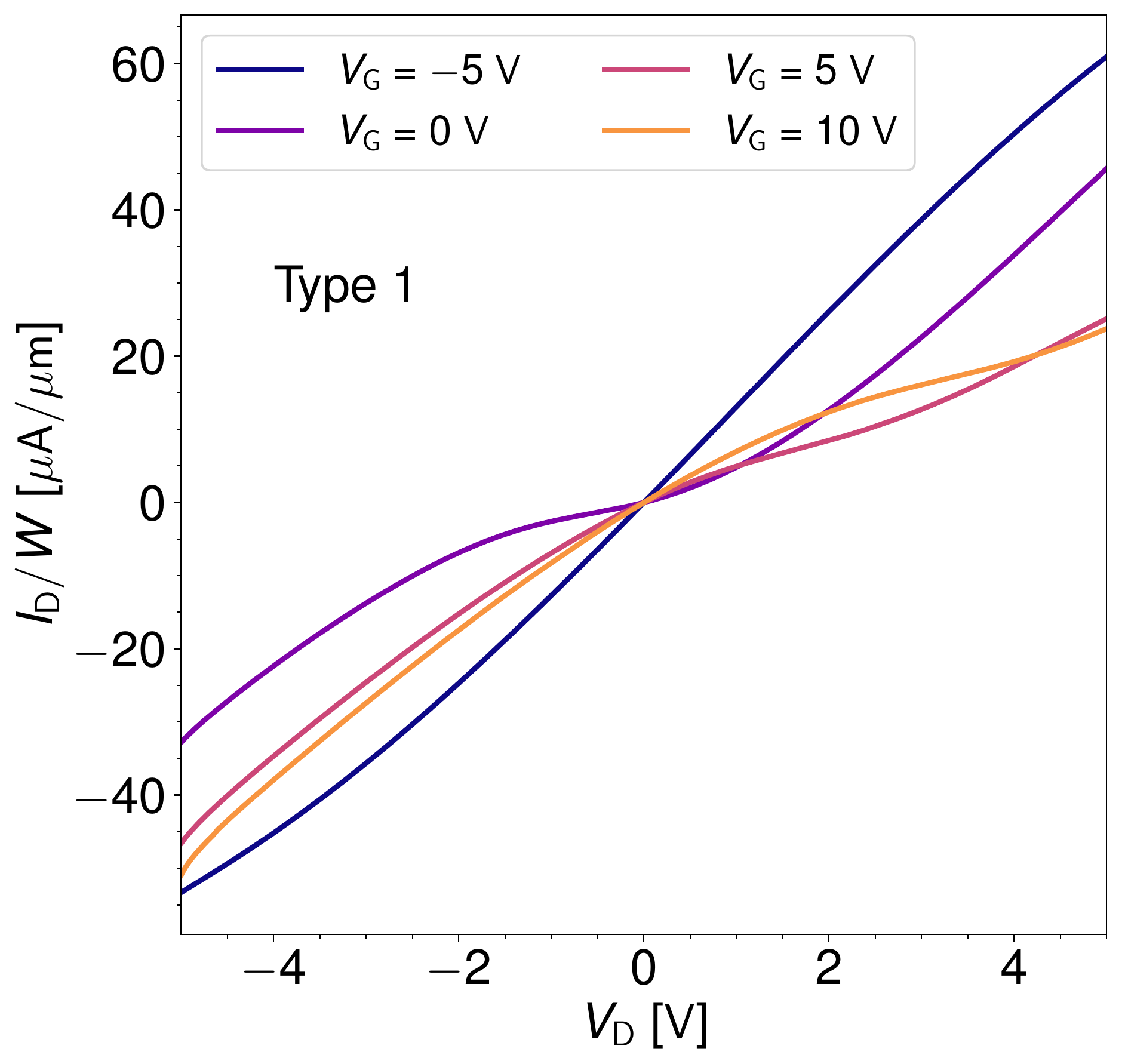}
        \vspace*{0.2cm}
        \label{fig:IdVd_1}
    \end{subfigure}
    \begin{subfigure}[c]{0.305\textwidth}
        \caption{}
        \centering
        \includegraphics[width=\textwidth]{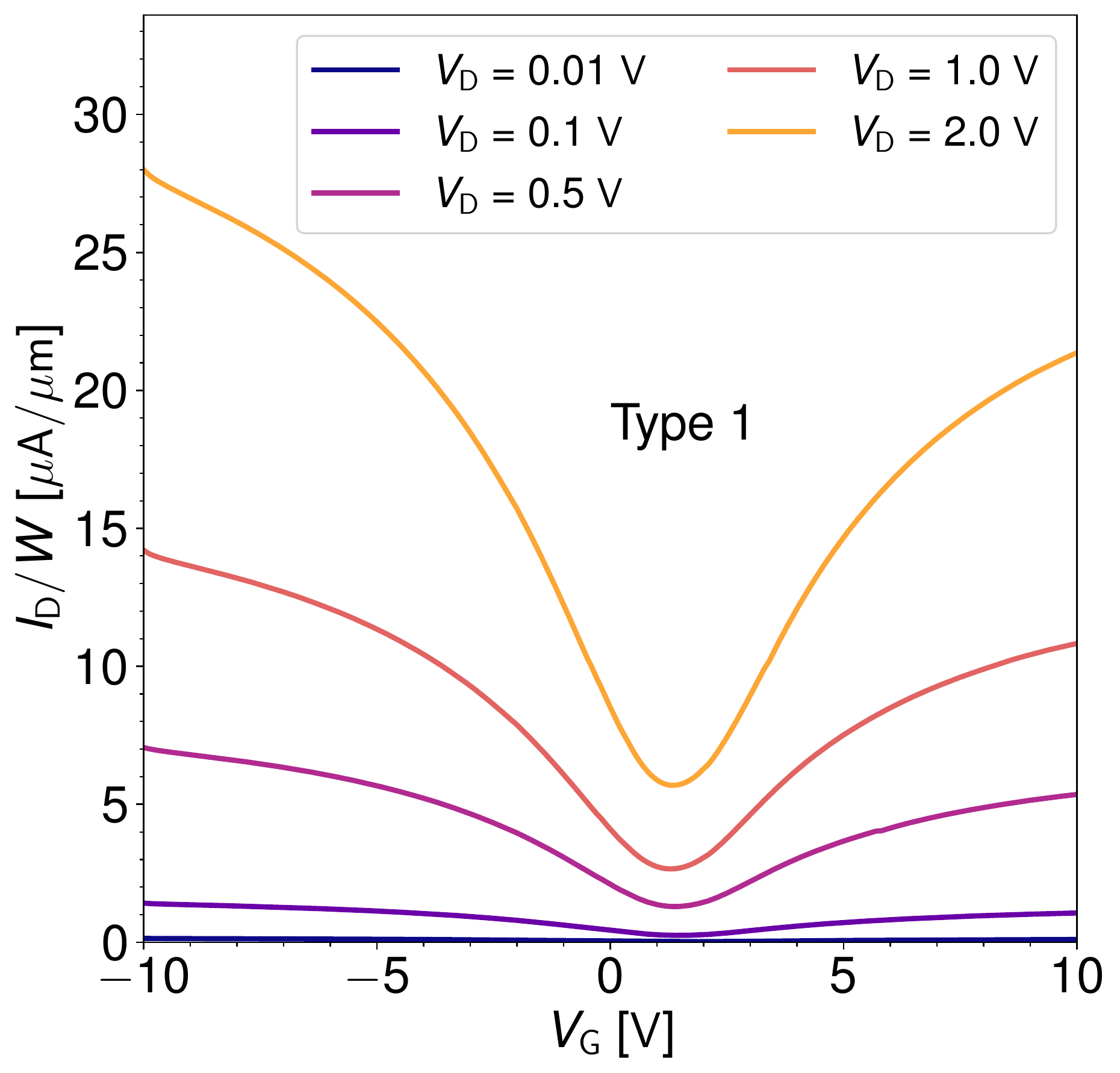} 
        \label{fig:IdVg_1}
    \end{subfigure}
    \begin{subfigure}[c]{0.305\textwidth}
        \caption{}
        \centering
        \includegraphics[width=\textwidth]{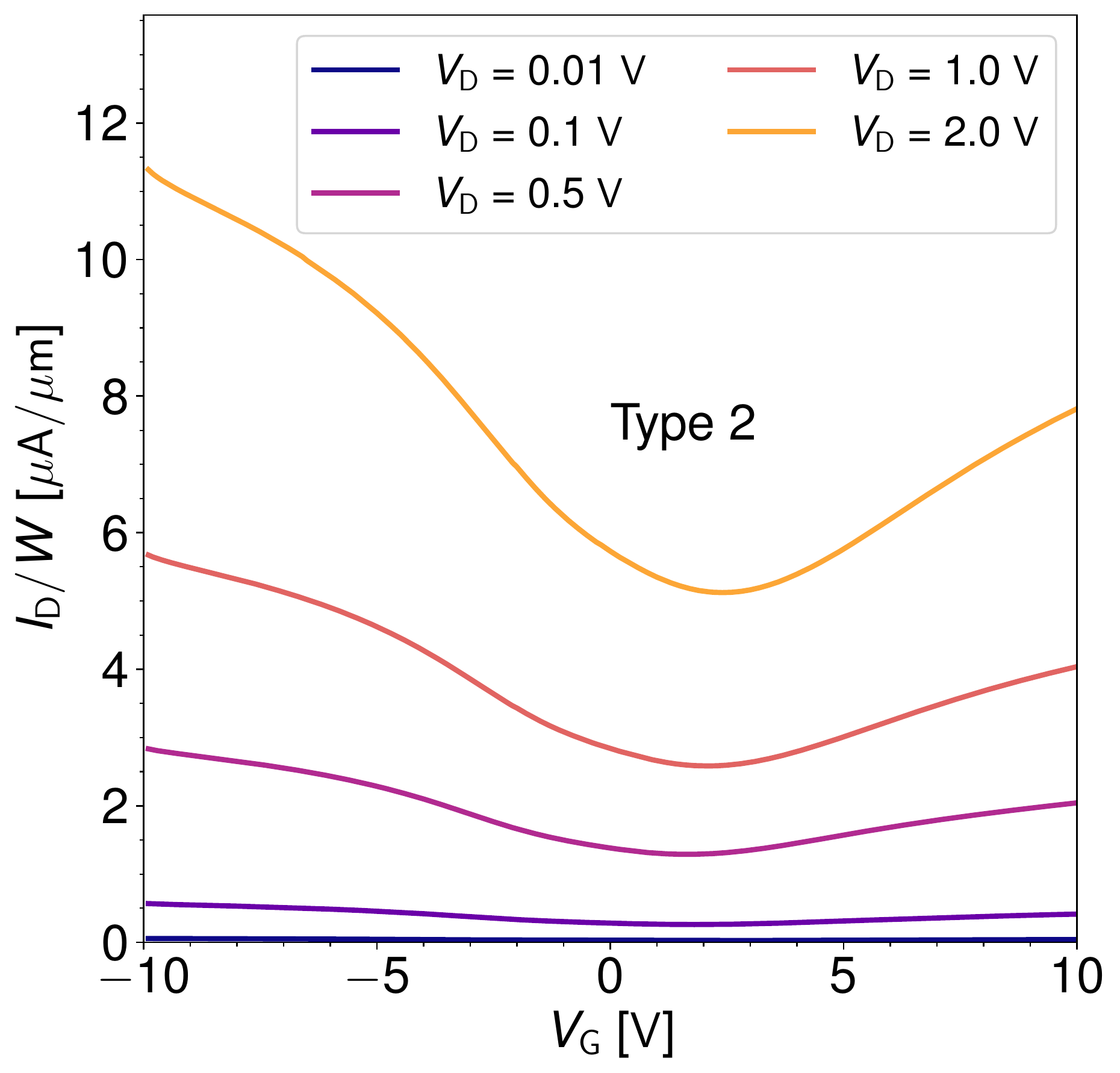} 
        \label{fig:IdVg_2}
    \end{subfigure}
    \begin{subfigure}[c]{0.31\textwidth}
        \caption{}
        \centering
        \includegraphics[width=\textwidth]{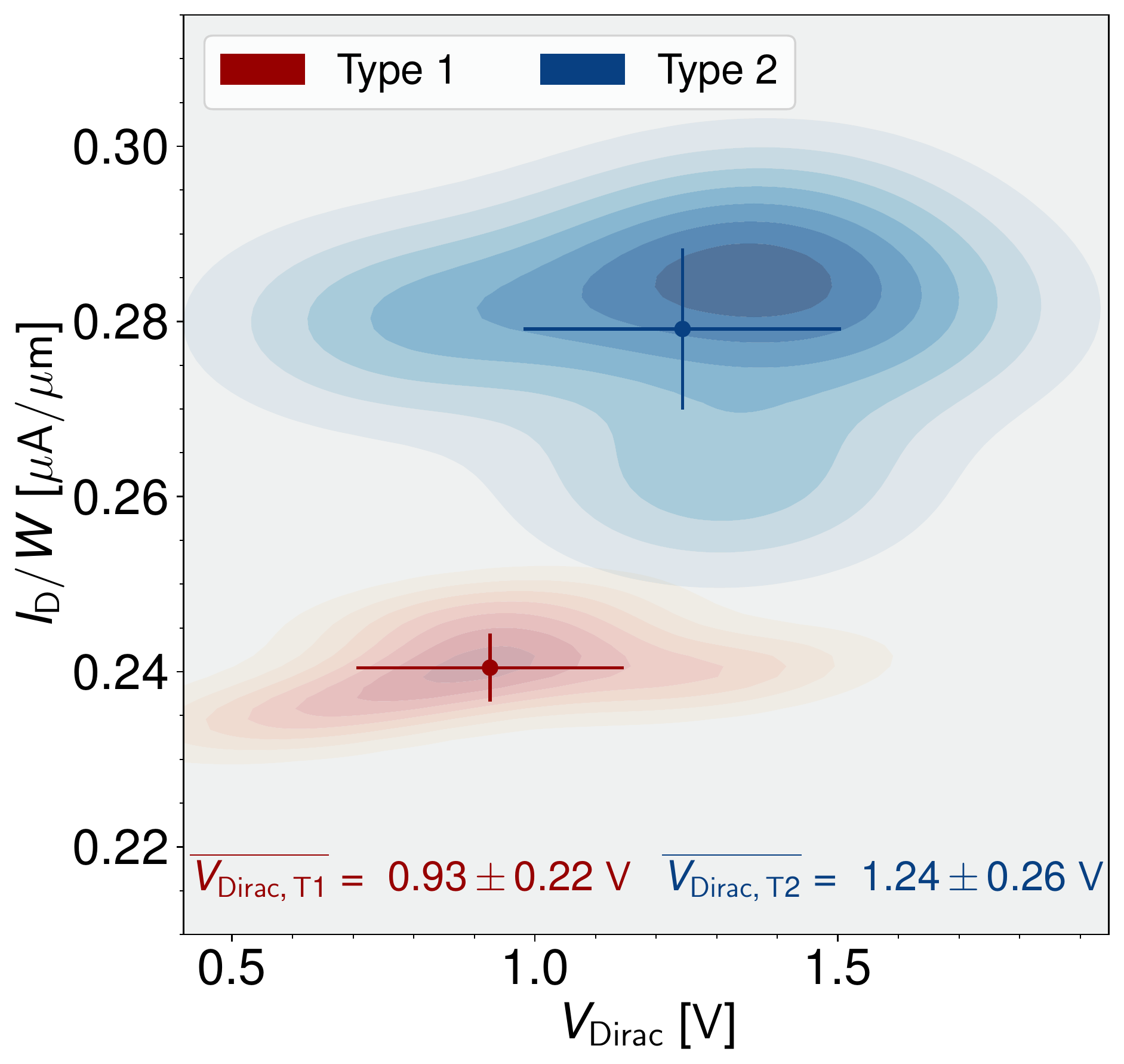} 
        \label{fig:Vd_comp}
    \end{subfigure}
\vspace*{-0.5cm}
\caption{\footnotesize
In (a) the schematic to the left shows the cross section while the optical microscope image to the right shows a top view of the device layout. The output ($I_{\mathrm{D}}$-$V_{\mathrm{D}}$) characteristics of a representative device of Type~1 are given in Figure (b) and the transfer ($I_{\mathrm{D}}$-$V_{\mathrm{G}}$) characteristics  in Figure (c). Type~2 GFETs have the same layout as Type~1 GFETs, but are based on a different CVD grown graphene layer from another vendor. In Figure (d) the transfer ($I_{\mathrm{D}}$-$V_{\mathrm{G}}$) characteristics of a representative device based on Type~2 graphene are shown. In Figure (e) the mean current and voltage at the Dirac point are compared for the two graphene types. These values were calculated from the transfer characteristics measured on 5 different GFETs of every type using $V_{\mathrm{D}}=\SI{0.1}{V}$ and a sweep time $t_{\mathrm{SW}}=\SI{1}{s}$. } 
\label{fig:GFETs}
\end{figure}

We examine GFETs fabricated on mechanically flexible polyimide~(PI) substrates\cite{Wang2019b} with graphene monolayers forming the channel with an area of $W\times L = \SI{100}{\micro m} \times \SI{160}{\micro m}$, see Figure \ref{fig:schematic}. In the top-gated device layout a \SI{40}{nm} thick, amorphous Al\textsubscript{2}O\textsubscript{3} layer grown by atomic layer deposition serves as gate oxide. We compare two nearly identical batches of GFETs with differently doped channel layers, as the graphene layers were purchased from different vendors using different parameters for the chemical vapor deposition process and the layer transfer. Type~1 graphene shows a smaller work function and in consequence a smaller distance of $E_\mathrm{F}$ to the Al\textsubscript{2}O\textsubscript{3} trap band ($\overline{E_\mathrm{T}}$). This small value of $\overline{E_\mathrm{T}} - E_\mathrm{F}$ predicts electrically unstable devices. In contrast, Type~2 graphene is p-doped with a higher distance of $E_\mathrm{F}$ to $\overline{E_\mathrm{T}}$, predicting electrically more stable FETs. In addition, the layer quality of Type~1 and Type~2 is different, revealing a higher concentration of defects in graphene in Type~2 graphene, for details see their respective Raman spectra in the SI, Figure S2.

To assess functionality and performance of our GFETs, the standard device characteristics, namely the output ($I_{\mathrm{D}}$-$V_{\mathrm{D}}$) and transfer ($I_{\mathrm{D}}$-$V_{\mathrm{G}}$) characteristics, are shown in Figures \ref{fig:IdVd_1} and \ref{fig:IdVg_1} for a representative Type~1 GFET. We observe ambipolar device operation with kinks in the output characteristics at higher $V_{\mathrm{D}}$, features typical for GFETs.~\cite{Schwierz2010} When comparing the transfer characteristics for Type~1 graphene with Type~2 graphene in Figure \ref{fig:IdVg_2}, it can be seen that the higher quality of Type~1 leads to higher current densities. Based on two-probe measurements of the $I_{\mathrm{D}}$-$V_{\mathrm{G}}$s we estimate the field-effect mobilities to reach up to \SI{5000}{cm^2/Vs} on Type~1 GFETs, four times higher than the average mobility of about \SI{600}{cm^2/Vs} on Type~2 GFETs. These results are expected based on the Raman analysis and originate from the higher amount of defects in Type~2 graphene. 
Negatively charged dopants in Type~2 lead to a higher variability and shift $V_\mathrm{Dirac}$ towards more positive voltages, see Figure \ref{fig:Vd_comp}. A more positive $V_\mathrm{Dirac}$ corresponds to a higher p-doping of the sample and correlates with a higher work function~($E_\mathrm{W}$)~\cite{Shi2010a, Kwon2012}. Pristine graphene has a work function of \SI{4.56}{eV}\cite{Yan2012a}, which is shifted towards higher values by p doping\cite{Shi2010a, Seo2014} and towards smaller values by n doping\cite{Park2011a, Kwon2012}. 
In order to estimate Fermi level location in the two graphene types, we approximate the charge carrier concentration $n$ at \SI{0}{V} gate bias based on the analytic approximation for the drain current of a MOSFET in the linear region
\begin{equation}
  n = \frac{I_\mathrm{D}\left(V_\mathrm{G} = \SI{0}{V}\right)}{\mu_\mathrm{eff} \frac{W}{L} q V_\mathrm{D}}.
\end{equation}
This expression gives a p-doping density for Type~1 graphene of $n_1 = \SI{8.2e11}{cm^{-2}}$ and for Type~2 graphene of $n_1 = \SI{4.7e12}{cm^{-2}}$, thus Type~2 graphene is more p-doped by an additional doping density of approximately $\SI{3.9e12}{cm^{-2}}$. These hole densities in the graphene layers at \SI{0}{V} gate voltage determine the work function via~\cite{Zhang2008a, Park2011a} 
\begin{equation}
 E_\mathrm{W} = \hbar \nu_\mathrm{F} \sqrt{\pi n}
\end{equation}
with  the Fermi velocity of graphene $\nu_\mathrm{F} = \SI{1.1e6}{ms^{-1}}$. In consequence, we estimate $E_\mathrm{W1}$ of Type~1 graphene to be \SI{4.7}{eV} and $E_\mathrm{W2}$ of Type~2 graphene to be \SI{0.2}{eV} higher at \SI{4.9}{eV}.

To further analyze our model system we fabricated devices with Type 1 graphene but using thermal SiO$_2$ on silicon and quartz substrates instead of the flexible PI layer. In addition, the quality of the interface between graphene and Al$_2$O$_3$ was modified by transferring single-layer CVD-grown hBN layers before the ALD deposition or by sputtering $\sim\SI{2}{nm}$ thick aluminum as a seed layer for the Al$_2$O$_3$ growth process. As will be seen in the discussion of these results in the SI Figure S3, the substrate primarily impacts the maximum current density whereas the quality of the interface with the Al$_2$O$_3$ impacts device stability.

\begin{figure}[htb!]
\vspace*{0.5cm}
\centering
\vspace*{-0.5cm}
    \begin{subfigure}{.59\textwidth}
    \caption{}
    \includegraphics[width=\textwidth]{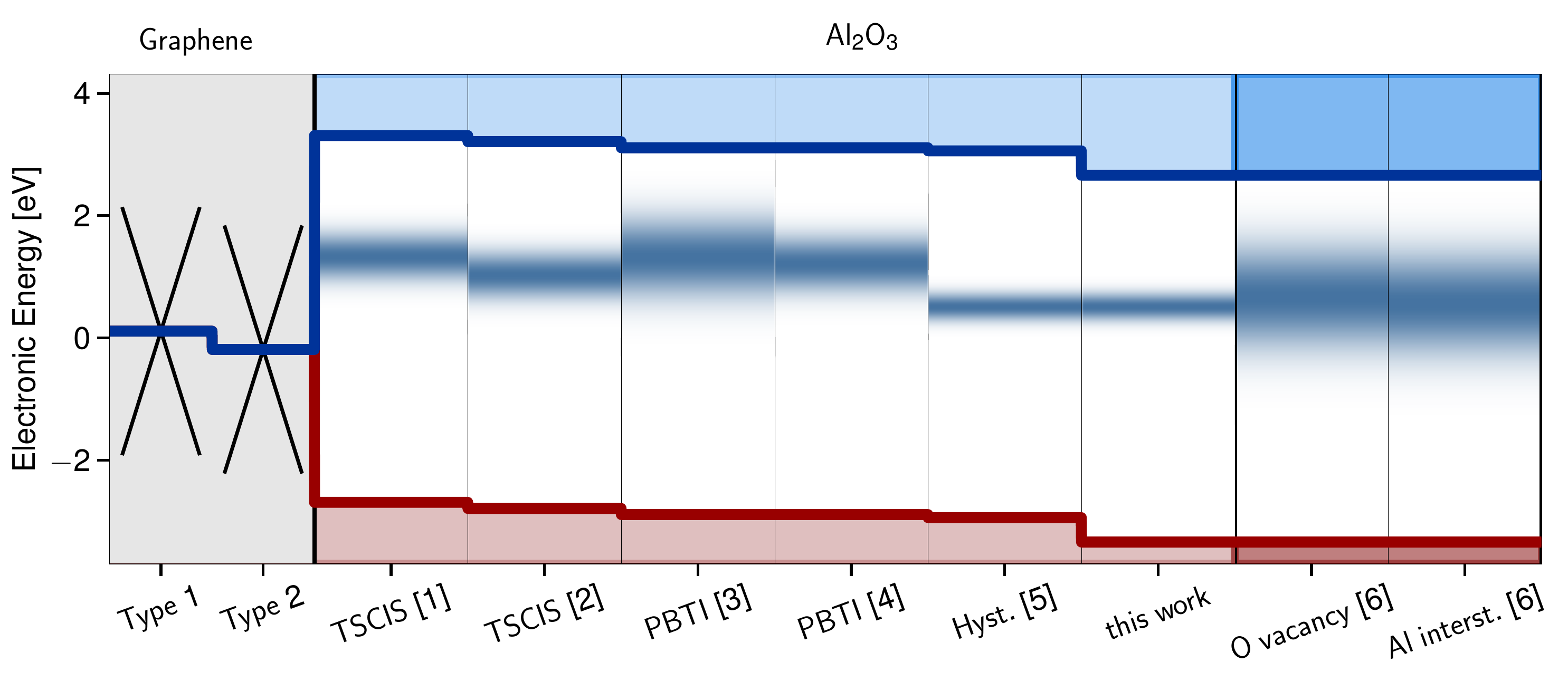}
    \label{fig:banddiagrams}
    \end{subfigure}
    \begin{subfigure}[c]{0.4\textwidth}
    \vspace*{-0.25cm}
    \caption{}
    \vspace*{0.25cm }
    \begin{minipage}{0.515\textwidth}
    \includegraphics[width=\textwidth]{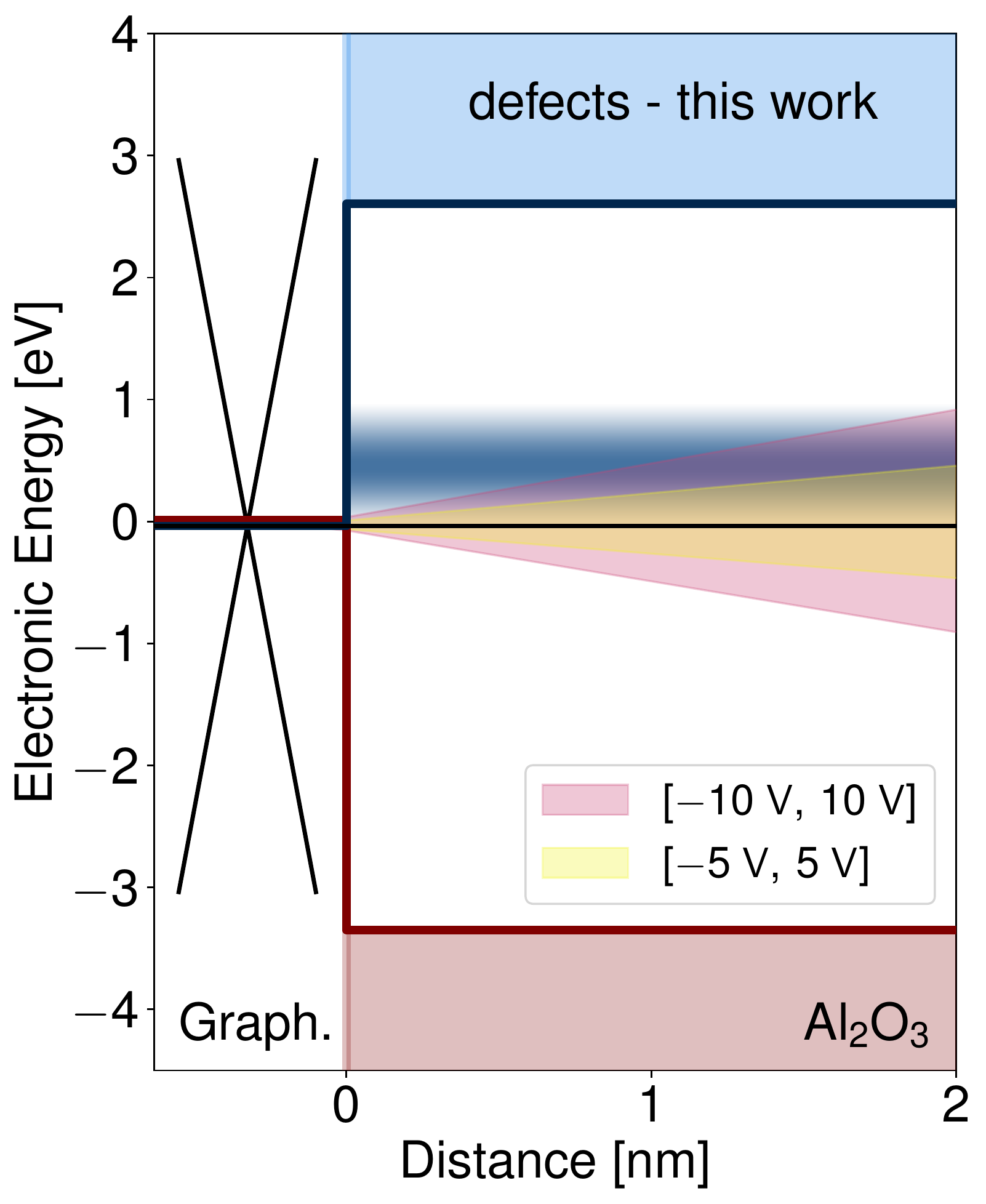}
    \end{minipage}
    \hspace*{-0.3cm}
    \begin{minipage}{0.485\textwidth}
    \includegraphics[width=\textwidth]{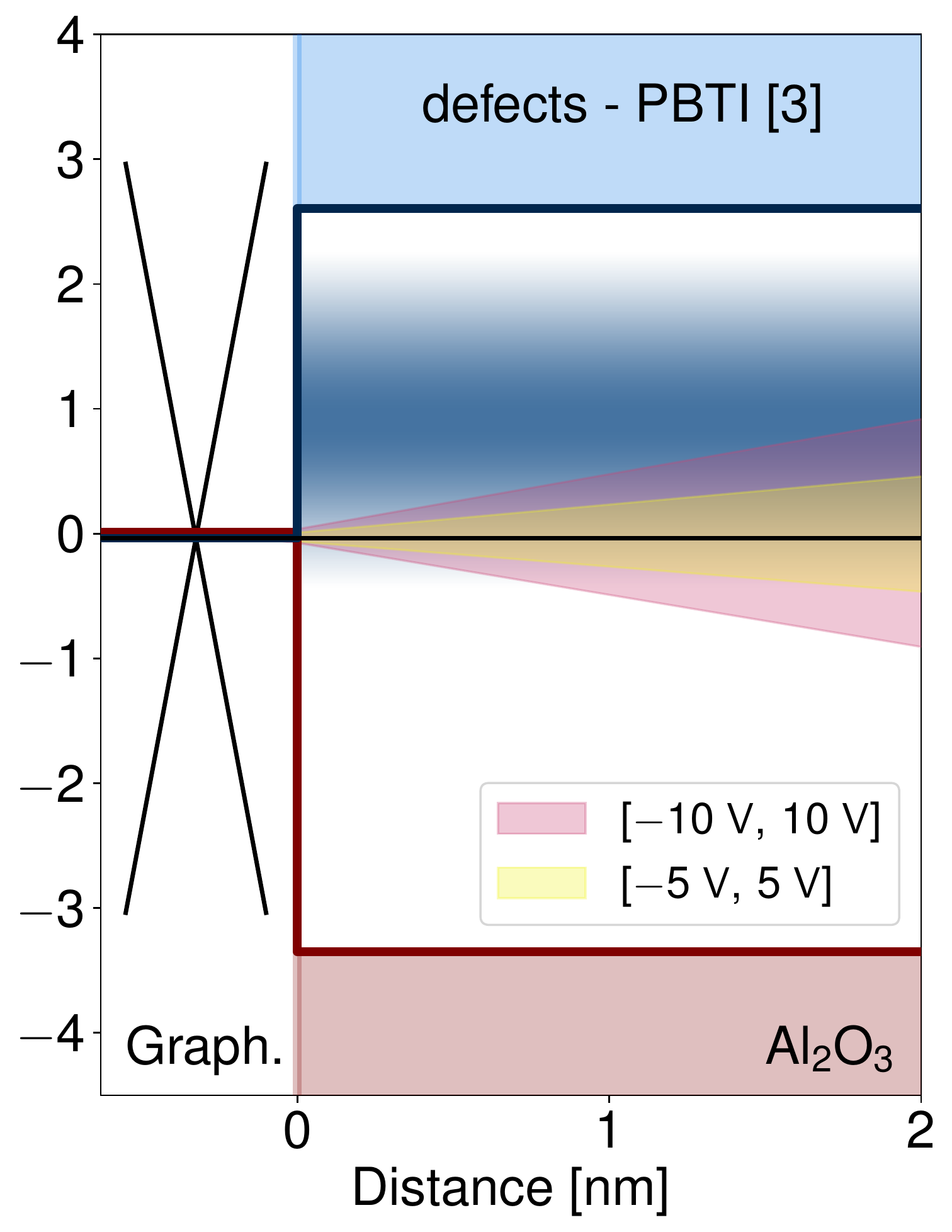}
    \end{minipage}
    \label{fig:active_region}
    \end{subfigure}
    \begin{subfigure}[c]{0.495\textwidth}
    \vspace*{-0.5cm}
    \hspace*{-0.5cm}
    \caption{Type 1.}
    \begin{minipage}{0.35\textwidth}
    \includegraphics[width=\textwidth]{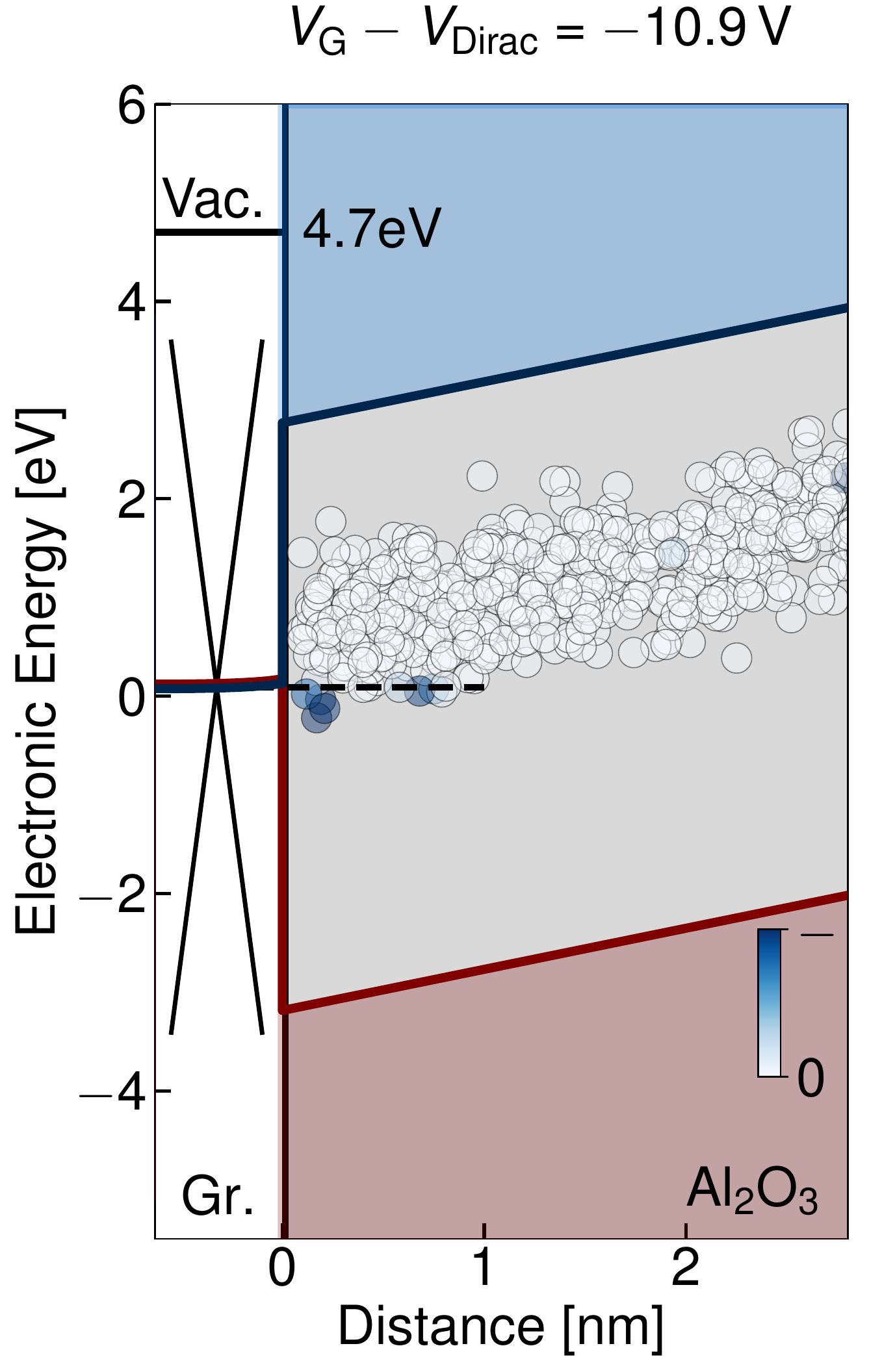}
    \end{minipage}
    \hspace*{-0.3cm}
    \begin{minipage}{0.33\textwidth}
    \includegraphics[width=\textwidth]{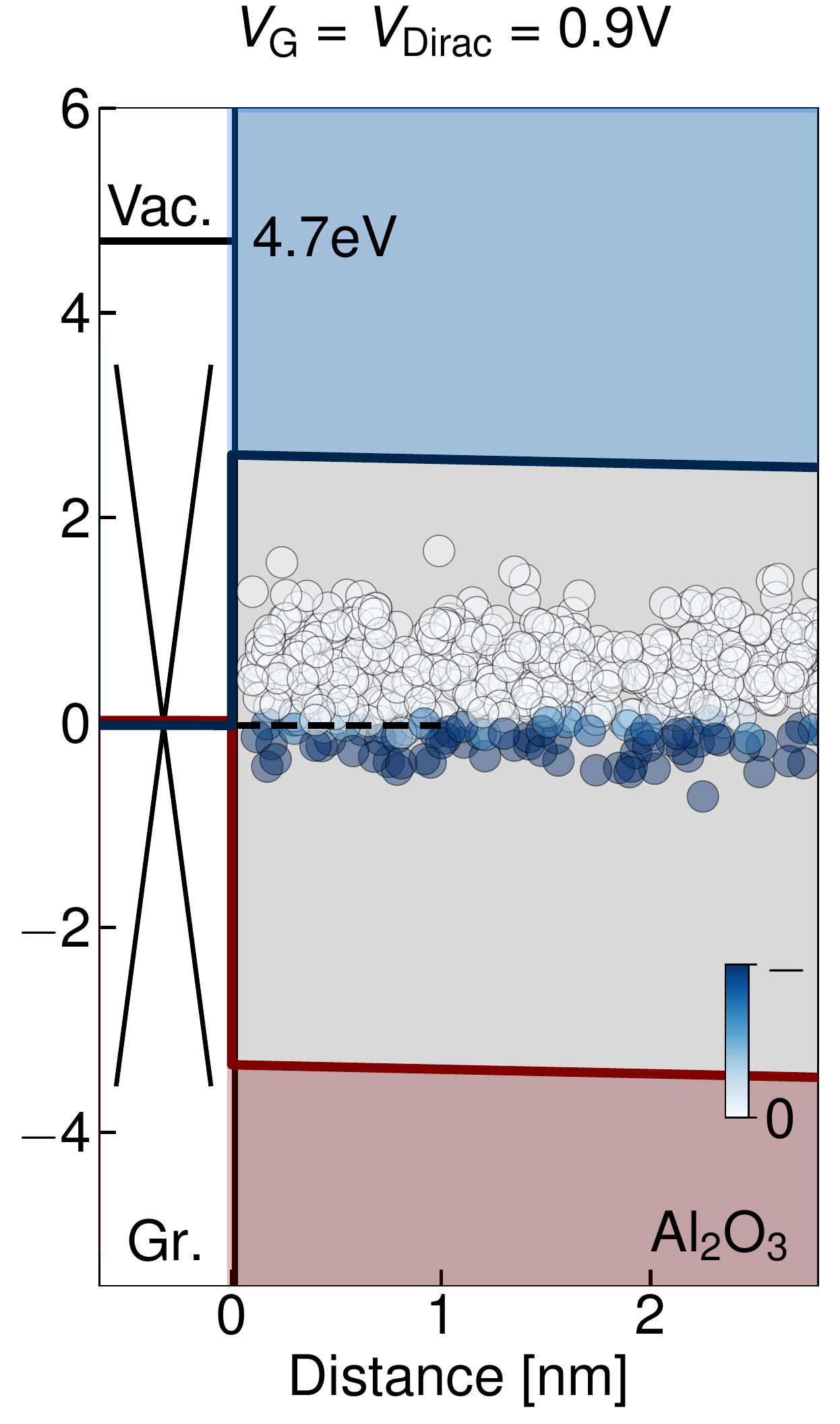}
    \end{minipage}
    \hspace*{-0.3cm}
    \begin{minipage}{0.33\textwidth}
    \includegraphics[width=\textwidth]{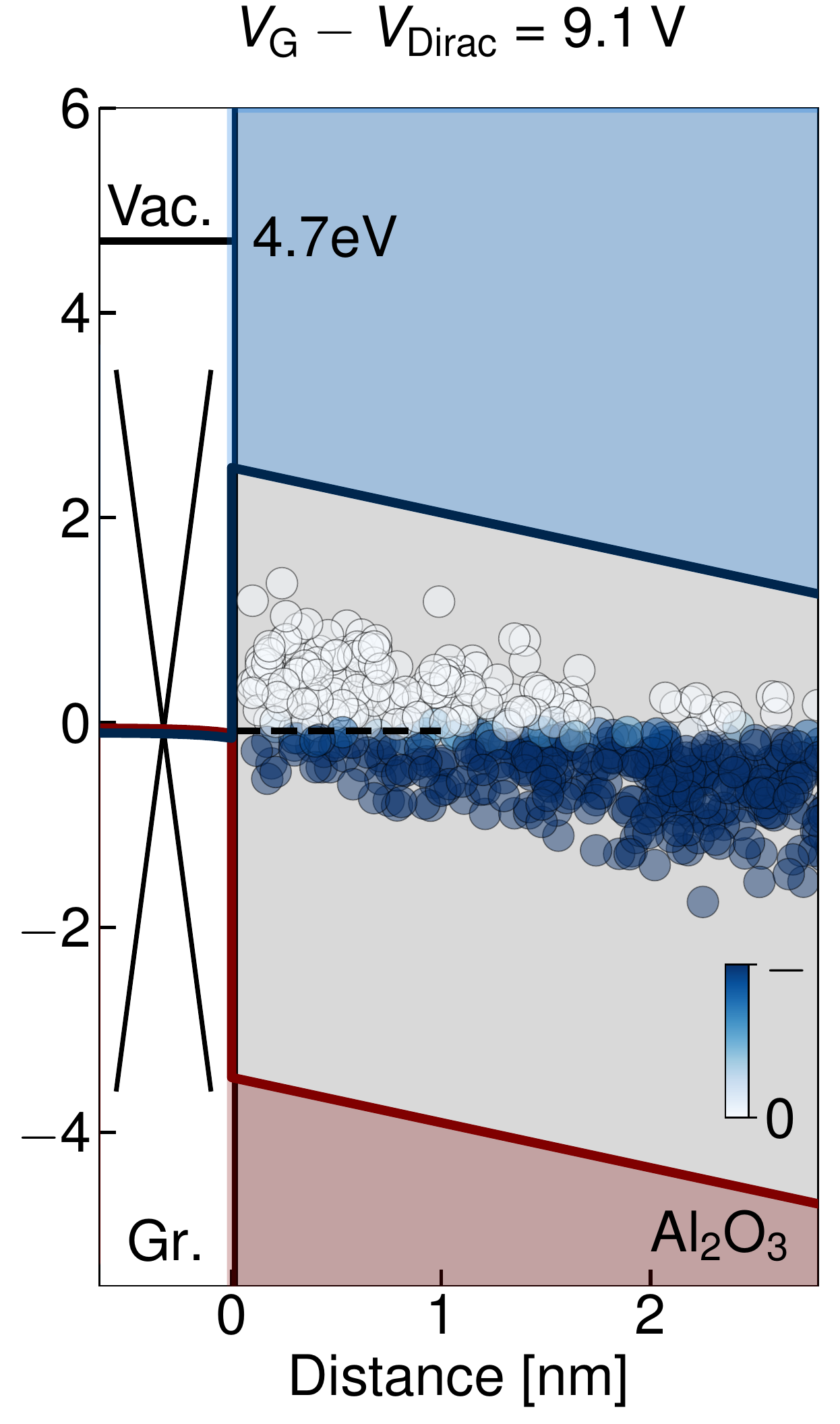}
    \end{minipage}
    \label{fig:bands_Type1}
    \end{subfigure}
    \begin{subfigure}[c]{0.49\textwidth}
    \vspace*{-0.5cm}
    \hspace*{-0.3cm}
    \caption{Type 2.}
    \begin{minipage}{0.33\textwidth}
    \includegraphics[width=\textwidth]{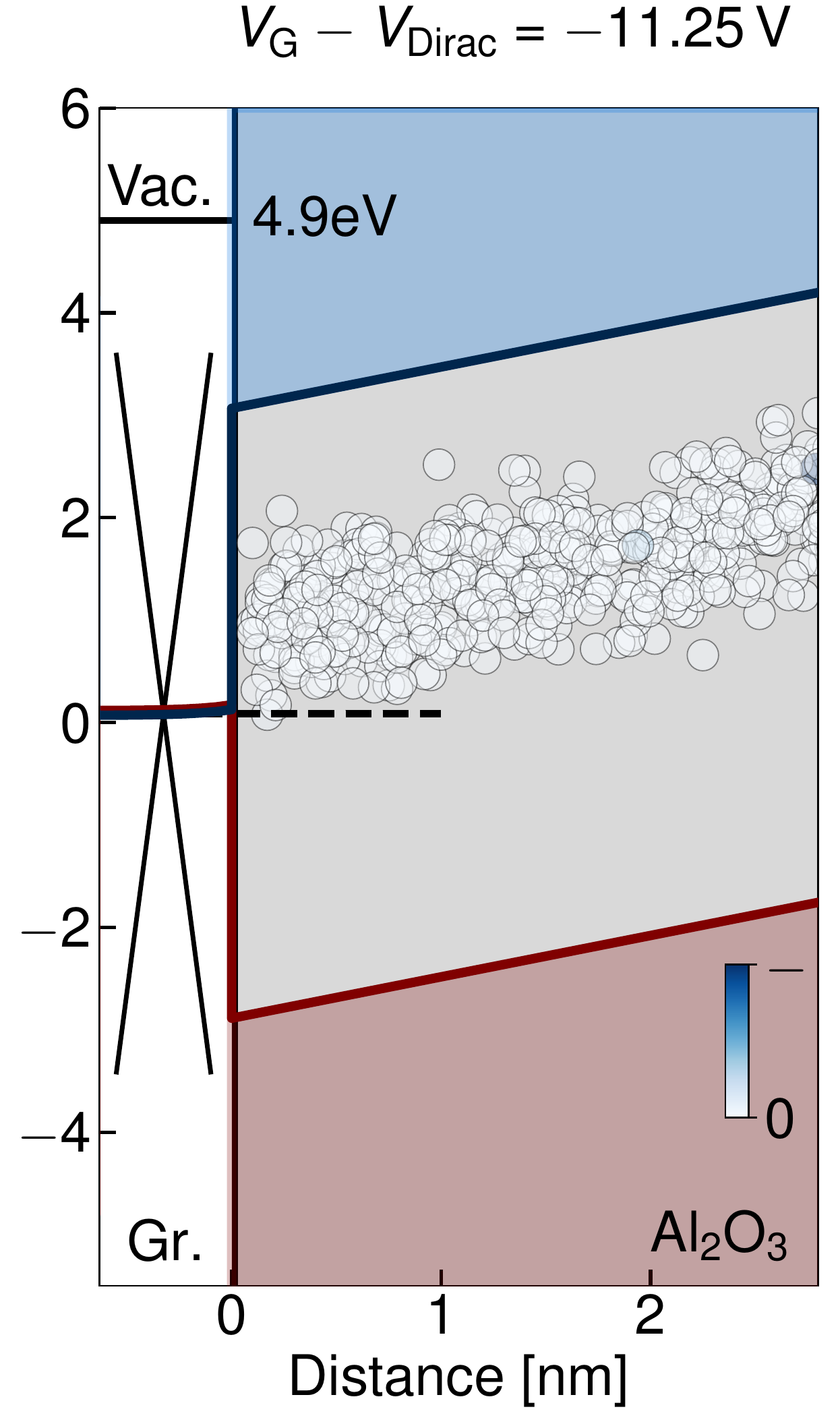}
    \end{minipage}
    \hspace*{-0.3cm}
    \begin{minipage}{0.33\textwidth}
    \includegraphics[width=\textwidth]{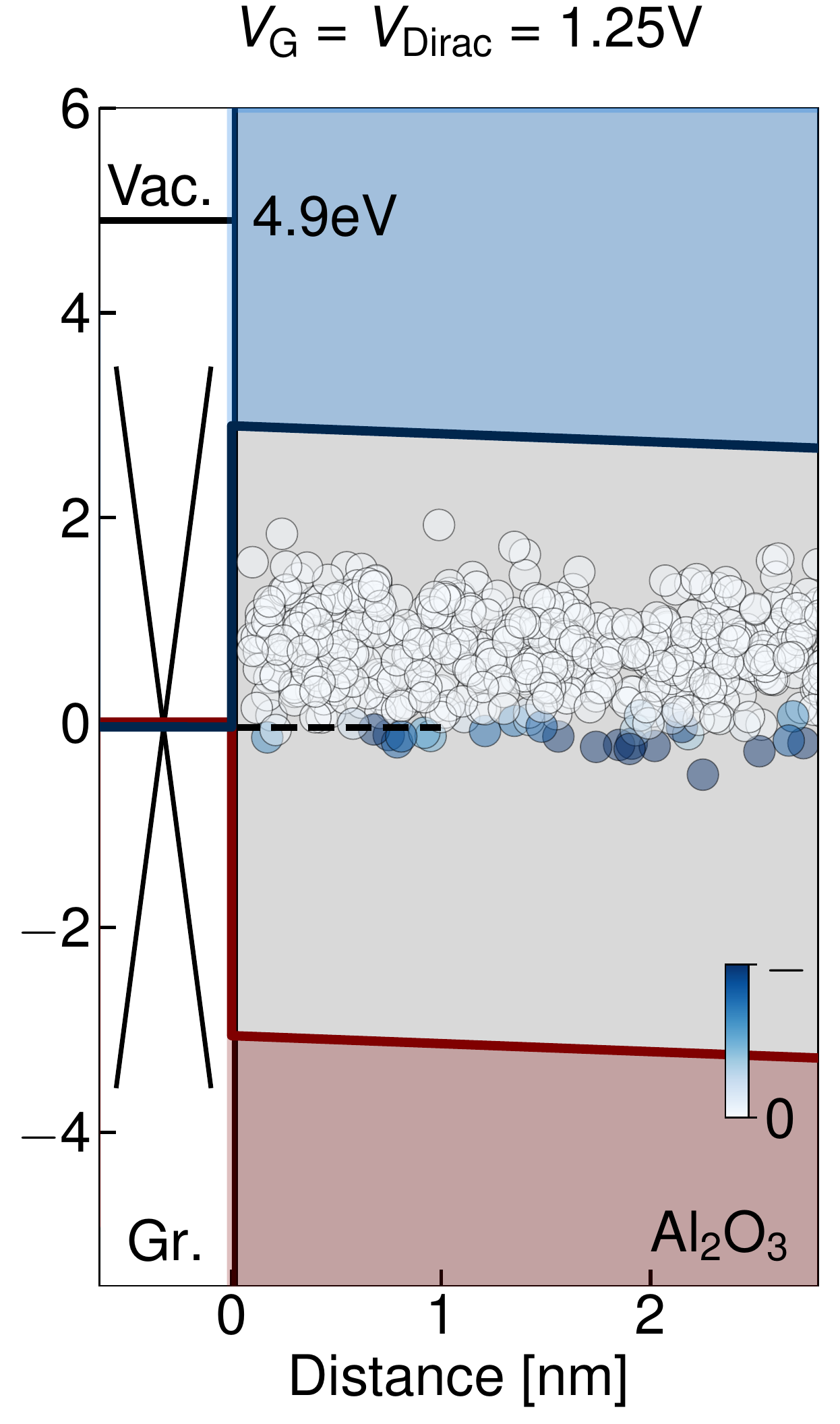}
    \end{minipage}
    \hspace*{-0.3cm}
    \begin{minipage}{0.33\textwidth}
    \includegraphics[width=\textwidth]{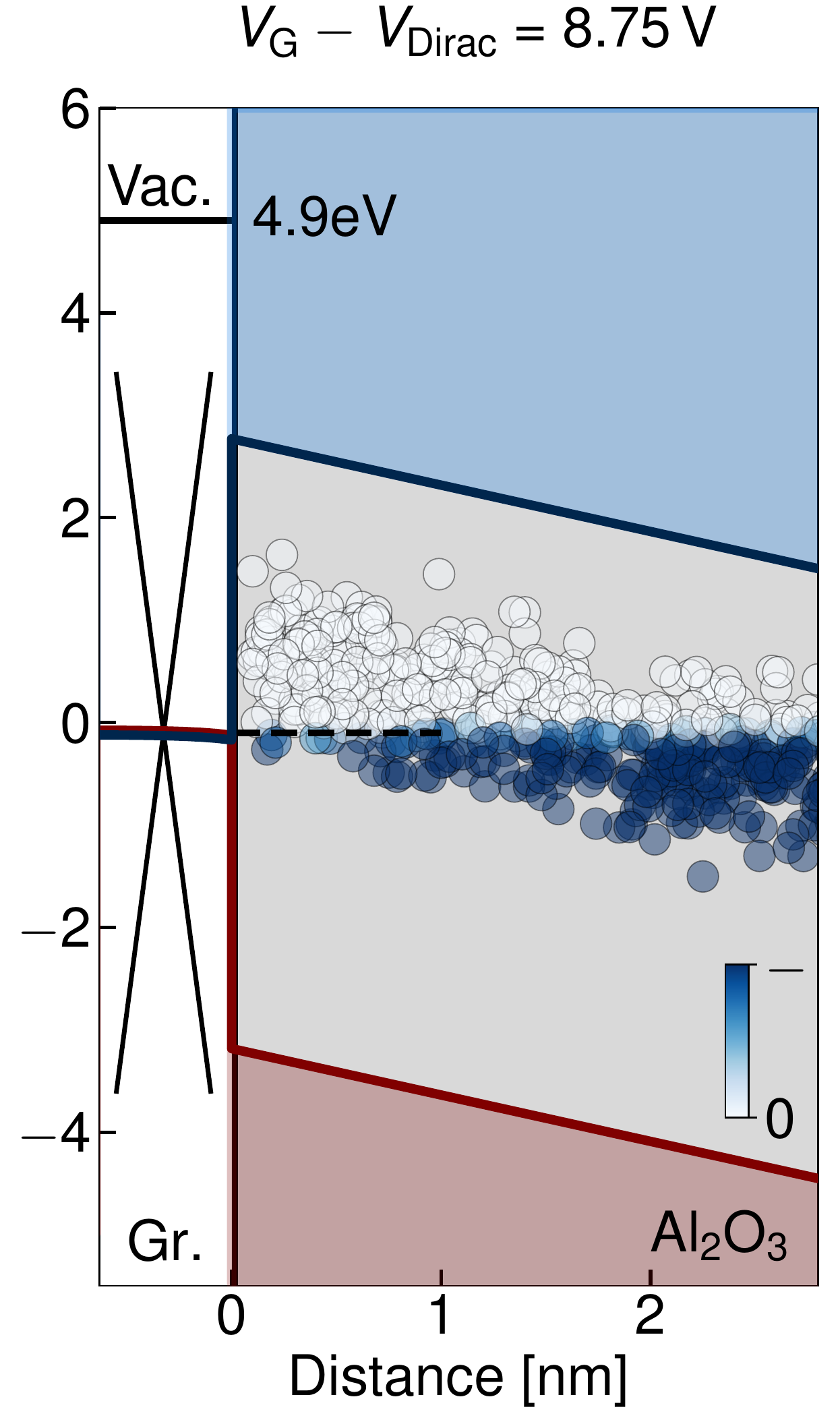}
    \end{minipage}
    \label{fig:bands_Type2}
    \end{subfigure}
\vspace*{0.25cm}
\caption{\footnotesize
In (a) the band diagram illustrates the alignment of the Al\textsubscript{2}O\textsubscript{3} defect band to the graphene of Type~1 and Type~2. 
To the left, the location of defect bands as extracted from experiments is shown:
[1]~\cite{Degraeve2008}, [2]~\cite{Zahid2010}, [3]~\cite{Franco2014}, [4]~\cite{Putcha2018}, [5]~\cite{ILLARIONOV17A}. To the right, the alignment of the defect band caused by oxygen vacancies and Al interstitials in amorphous Al\textsubscript{2}O\textsubscript{3} is shown according to the DFT calculations presented by Dicks~\textit{et al.} [6]~\cite{DICKS19}.
In (b) the active region which is probed by measurements in the $[\SI{-5}{V},\,\SI{5}{V}]$ and $[\SI{-10}{V},\,\SI{10}{V}]$ range is shown for two defect band alignments for Type~1 GFETs.
In (c) schematic band diagrams show the charging and discharging of defects in Al\textsubscript{2}O\textsubscript{3} for Type 1 graphene with a work function of $E_\mathrm{W}=\SI{4.7}{eV}$.
In (d) the band diagrams for Type 2 graphene with $E_\mathrm{W}=\SI{4.9}{eV}$ are shown.}
\label{fig:Al2O3}
\end{figure}

To accurately determine the alignment of $E_\mathrm{F}$ in graphene to the electron trapping band of the amorphous Al\textsubscript{2}O\textsubscript{3} gate oxide at $\overline{E_\mathrm{T}}$, the precise location of the oxide defect band is essential. 
Several studies have investigated the alignment of this defect band using trap spectroscopy by charge injection and sensing (TSCIS)~\cite{Degraeve2008,Zahid2010}, BTI~\cite{Franco2014, Putcha2018} and hysteresis measurements~\cite{ILLARIONOV17A}.  Obtained defect band alignments of Al\textsubscript{2}O\textsubscript{3} from literature are shown in Figure \ref{fig:banddiagrams}, with parameters listed in Table S1 in the SI. 
Based on density functional theory (DFT) calculations, this defect band can be associated with either oxygen vacancies~\cite{Guo2016a, DICKS19} or aluminum interstitials~\cite{DICKS19}. We use for our study a normally distributed defect band with the mean defect level being located at $E_\mathrm{C} - E_\mathrm{T} = 2.15\pm$\SI{0.3}{eV} below the conduction band edge of Al\textsubscript{2}O\textsubscript{3}. The electron affinity ($\chi$) of Al\textsubscript{2}O\textsubscript{3}, which determines the location of the conduction band edge, varies in literature. Here, we use \SI{1.96}{eV} as obtained from internal photoemission measurements~\cite{Afanasev2003}. We use the same level as in~\cite{ILLARIONOV17A} of \SI{4.1}{eV} which is below the level of \SI{3.4}{eV} obtained in most other experimental references~\cite{Zahid2010, Putcha2018}. However, the theoretical value of \SI{4}{eV}~\cite{DICKS19} is closer to our estimate of the defect bands' location. In addition, for all measurement ranges used in our work, we probe only the lower part of a potentially wider defect band further up, as measured by other methods~\cite{Franco2014}. This is illustrated in Figure \ref{fig:active_region}, where the regions which can be probed by measurements are shaded in light red and yellow. These shaded regions reach the upper edge of the defect band used in this work but cover only the lower part of the wider defect band reported by Franco \textit{et al.}~\cite{Franco2014}. For Type~1 graphene $E_\mathrm{F}$ is aligned within the defect band~(small $\overline{E_\mathrm{T}} - E_\mathrm{F}$, electrically unstable), see Figure \ref{fig:bands_Type1}, whereas it is aligned below the defect band for Type~2 graphene~(high $\overline{E_\mathrm{T}} - E_\mathrm{F}$, electrically stable), see Figure~\ref{fig:bands_Type2}. We will see in the next sections that the \SI{200}{meV} downwards shift of the Fermi level of Type~2 graphene is sufficient to render the $V_\mathrm{Dirac}$ of these GFETs more stable.


\section{Hysteresis dynamics}
\label{sec:hyst}

\begin{figure}[!htb]
\centering
    \begin{subfigure}[c]{0.56\textwidth}
    \caption{}
    \centering
    \includegraphics[width=\textwidth]{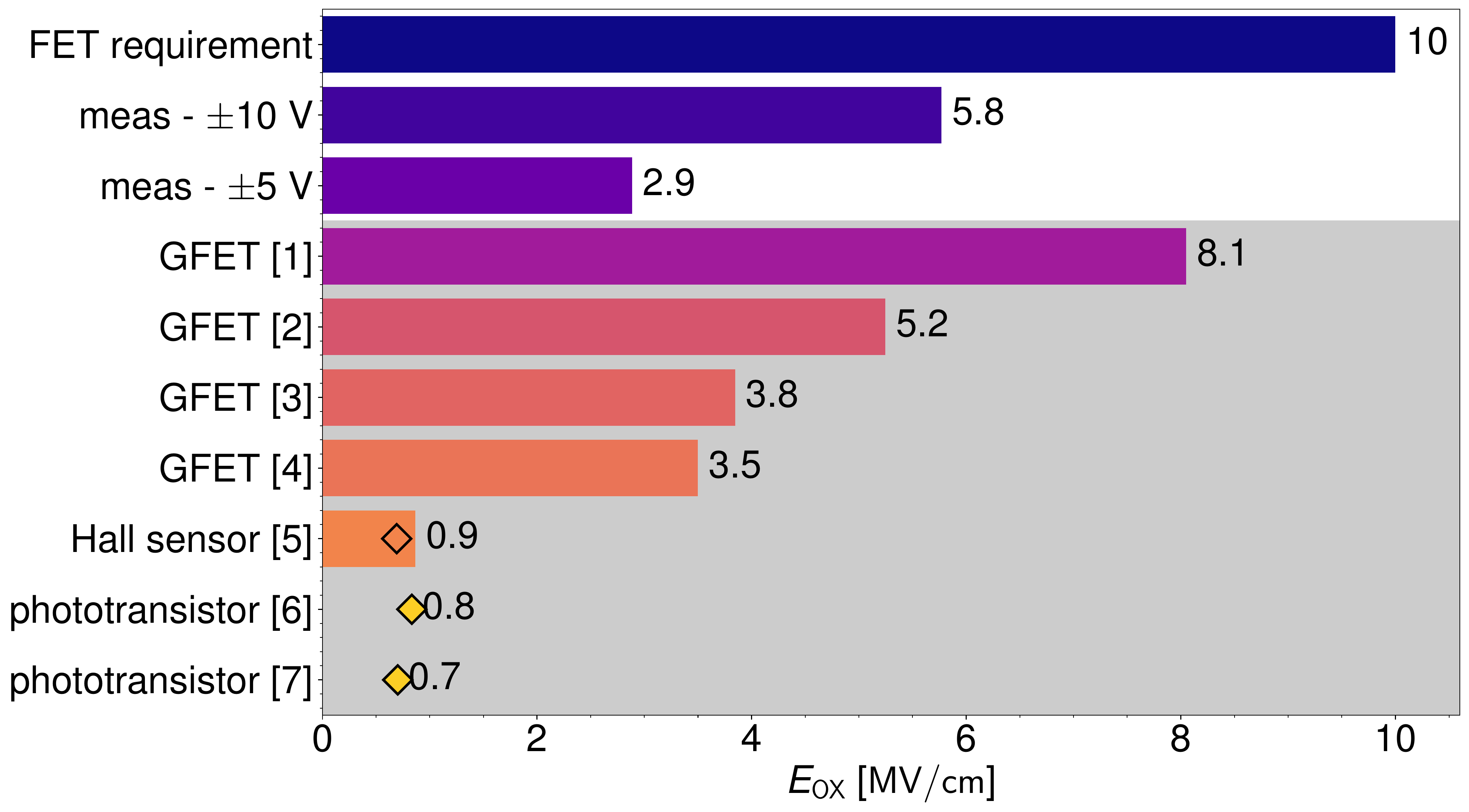}
    \label{fig:operate}
    \end{subfigure}
    \hspace*{1.15cm}
    \begin{subfigure}[c]{0.33\textwidth}
    \caption{}
    \includegraphics[width=\textwidth]{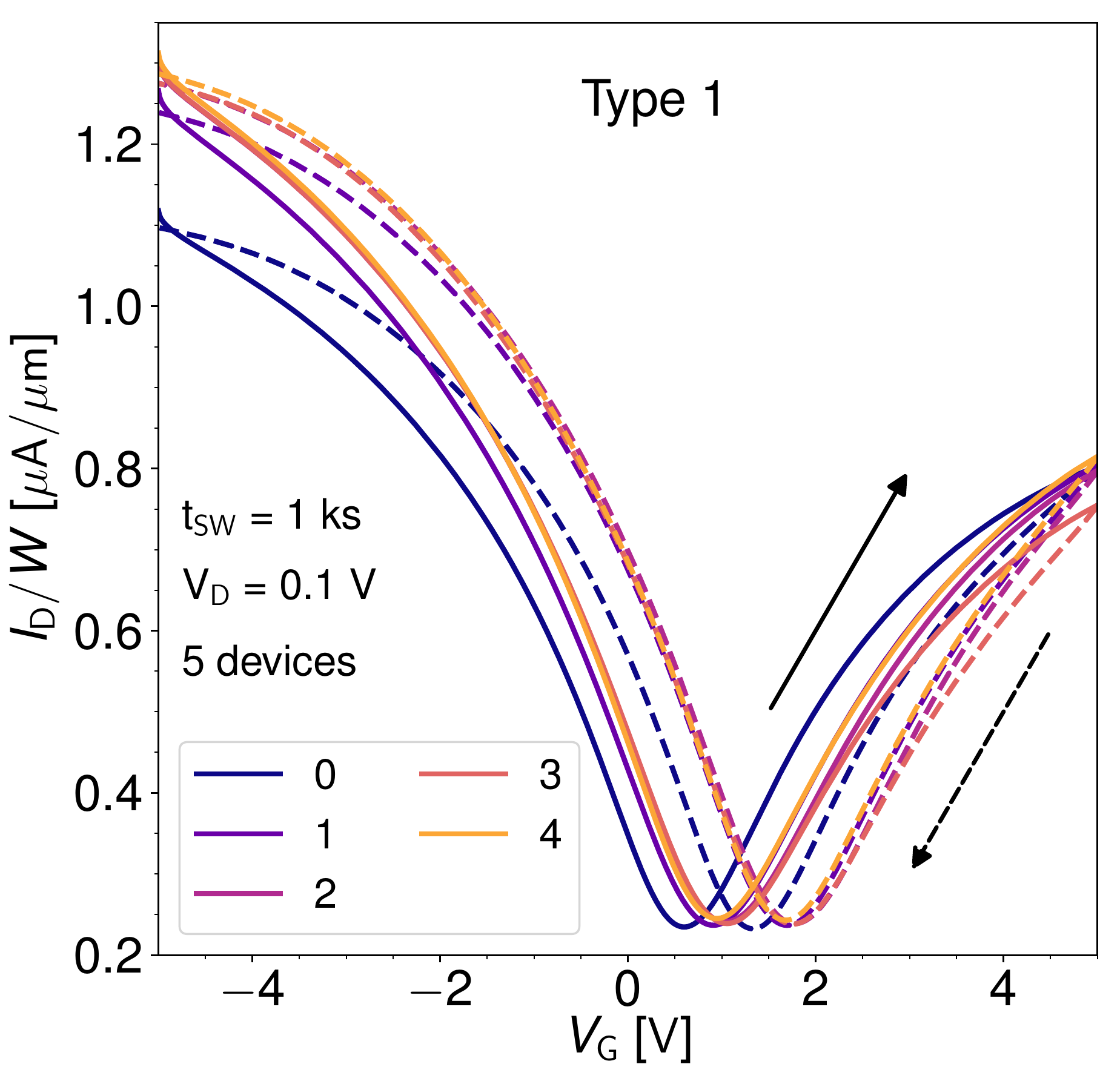}
    \label{fig:hyst_1_comp}
    \end{subfigure}\\
    \begin{subfigure}[c]{0.33\textwidth}
    \vspace*{-0.4cm}
    \caption{}
    \includegraphics[width=\textwidth]{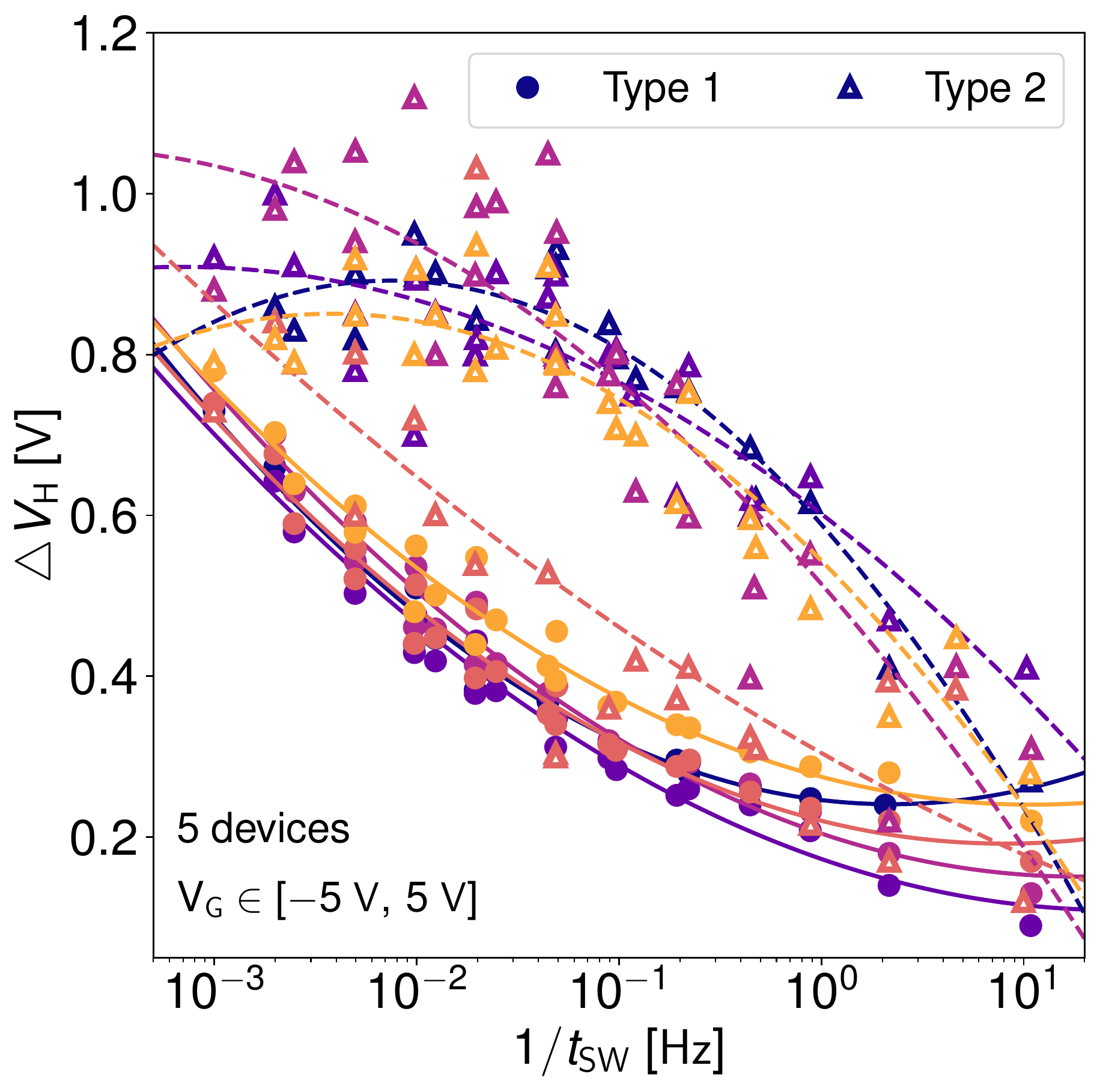}
    \label{fig:hyst_freq_5}
    \end{subfigure}
    \begin{subfigure}[c]{0.33\textwidth}
    \vspace*{-0.4cm}
    \caption{}
    \includegraphics[width=\textwidth]{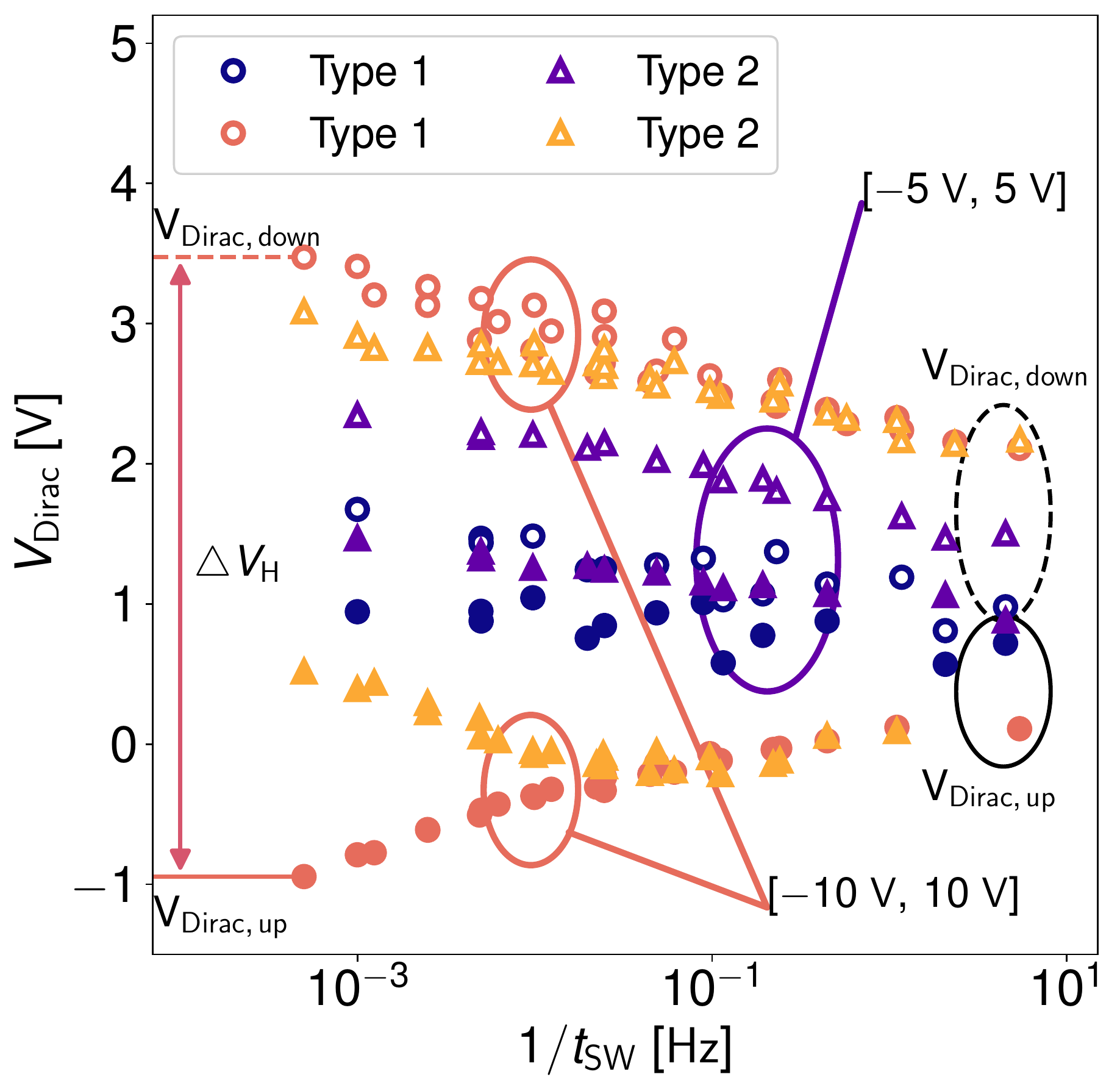}
    \label{fig:hyst_Dirac_10}
    \end{subfigure}
    \begin{subfigure}[c]{0.32\textwidth}
    \vspace*{-0.4cm}
    \caption{}
    \includegraphics[width=\textwidth]{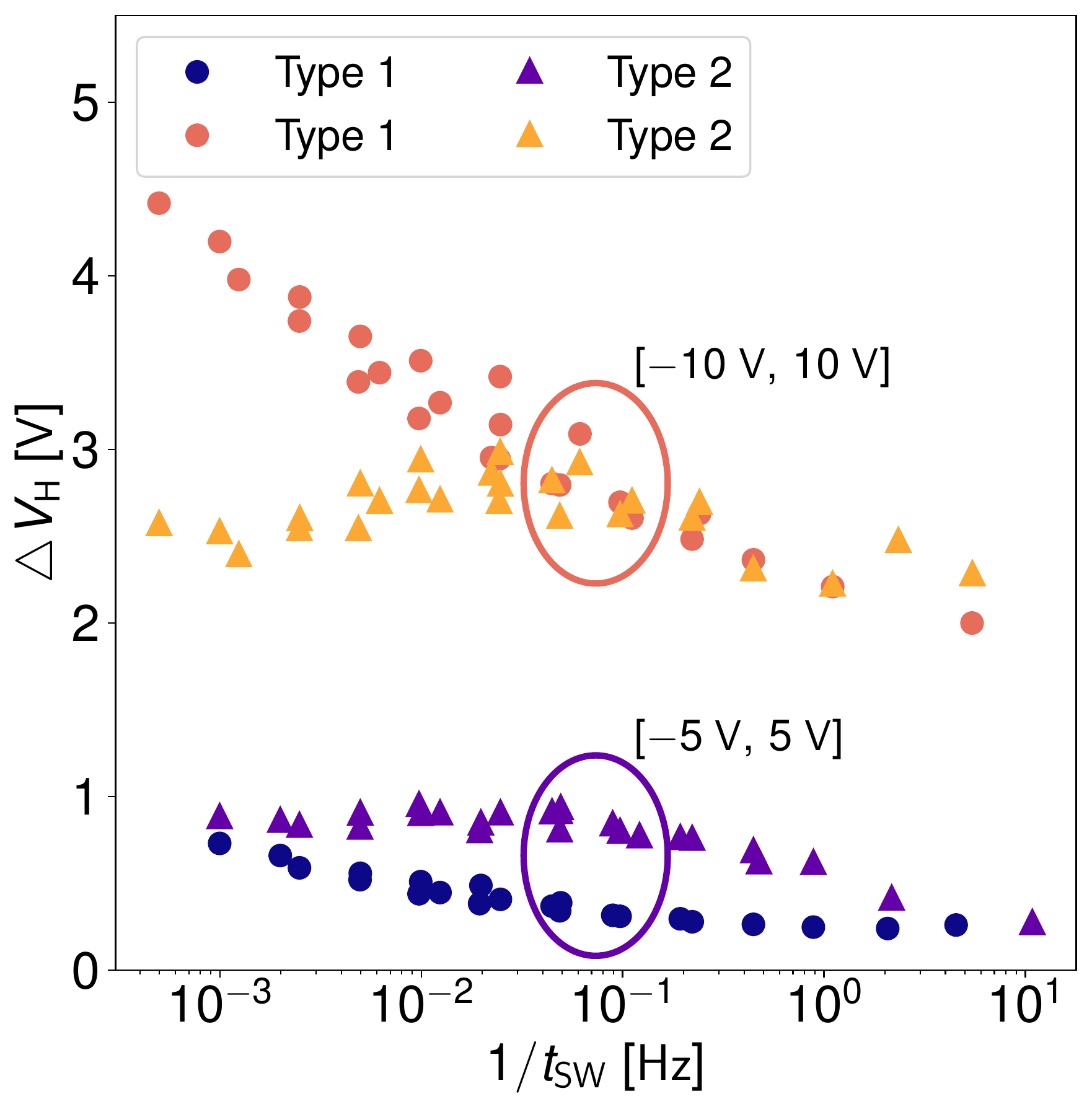}
    \label{fig:hyst_width_10}
    \end{subfigure}
\vspace*{-0.6cm}
 \caption{\footnotesize 
 In (a) the electrical oxide fields for measured voltage ranges are compared to voltage ranges for various applications. The field up to which GFETs should show stable operation is shown at the top in dark blue~\cite{IRDS2020} and the ranges we use here are shown in purple below. Application scenarios are from [1]~\cite{Wei2016}, [2]~\cite{Bonmann2019}, [3]~\cite{Kedzierski2009}, [4]~\cite{Lemme2007}, [5]~\cite{Uzlu2019}, [6]~\cite{Mueller2010}, [7]~\cite{KonSTANTATOS12}. 
In (b) the hysteresis in the transfer characteristic measured on 5 different GFETs is shown, illustrating the variability of the devices.
 In (c) the hysteresis width as a function of $1/t_{\mathrm{SW}}$ is shown for 5 different devices for each graphene type. Full circles stand for Type~1 GFETs and empty triangles for Type~2 GFETs, the solid and dashed lines are guides to the eye for Type~1 and Type~2 respectively. 
 In (e) the Dirac point shifts of the up and down sweep are shown with empty/full symbols for $V_\mathrm{Dirac,down}$/$V_\mathrm{Dirac,up}$. 
 In (f) the hysteresis width for Type~1 and Type~2 is shown.
 }
 \label{fig:hyst1}
\end{figure}

As the observed hysteresis depends critically on the voltage ranges used for the gate voltage sweeps, we compare
the bias ranges used with ranges for various applications in Figure~\ref{fig:operate}.
State-of-the-art silicon transistors operate at an electric gate field of \SI{10}{MV/cm} estimated based on the equivalent oxide thickness~(EOT)~\cite{IRDS2020}, up to which logical switches should show stable operation. As our devices employ an aluminum oxide layer of \SI{40}{nm} physical thickness as a gate oxide, their EOT amounts to $\sim$\SI{17}{nm}. When used in radio-frequency~(RF) circuits the electric gate fields span from \SI{3.5}{MV/cm} to \SI{8.1}{MV/cm}~\cite{Lemme2007, Kedzierski2009, Wei2016, Bonmann2019} and thus the gate oxide fields we investigate here are standard operation conditions for RF applications. If on the other hand GFETs are used as sensors in the form of phototransistors~\cite{Mueller2010, KonSTANTATOS12} or Hall elements~\cite{Uzlu2019}, moderate gate fields of up to \SI{1}{MV/cm} are sufficient to maximize responsivity.

As a first step in the hysteresis evaluation on our prototypes, we compare the double sweep transfer characteristics for a small voltage range of $\left[\SI{-5}{V}, \SI{5}{V}\right]$ on five GFETs based on Type~1 graphene in Figure \ref{fig:hyst_1_comp}. We see little variability which is confirmed when studying the hysteresis width $\Delta V_{\mathrm{H}}$ as a function of the inverse sweep time, the sweep frequency ($f=1/t_\mathrm{SW}$). In Figure \ref{fig:hyst_freq_5} the hysteresis width as a function of the sweep frequency is shown for five GFETs based on Type~1 and five GFETs based on Type~2. Type~2 devices show a considerably higher variability of $\Delta V_{\mathrm{H}}$ than Type~1 devices, which is linked to the increased variability of $V_{\mathrm{Dirac}}$ on Type~2, see Figure \ref{fig:Vd_comp}.
In addition, on Type~2 GFETs the hysteresis is higher and for both types the largest hysteresis is observed for the slowest sweeps as there the largest number of oxide defects can change their charge state. An increased bias range of $\left[\SI{-10}{V},\SI{10}{V}\right]$ increases the hysteresis, because more oxide defects become accessible for charge transfer, as can be seen in Figure \ref{fig:hyst_width_10}. To shed more light on this behavior, the dynamics of the Dirac voltage shifts are analyzed as a function of the sweep frequency in Figure \ref{fig:hyst_Dirac_10}. For the $\left[\SI{-5}{V},\SI{5}{V}\right]$ sweep, $V_{\mathrm{Dirac, up}}$ and $V_{\mathrm{Dirac, down}}$ as a function of the sweep frequency show similar slopes for both types. However, for the \SI{10}{V} sweep range and Type~1, $V_\mathrm{Dirac,up}$ is shifted to more negative voltages in slow sweeps, while $V_\mathrm{Dirac,down}$ is shifted to more positive voltages. This indicates that for large sweep ranges on Type~1 GFETs, a significant amount of electrons are emitted from oxide traps between \SI{-10}{V} and $V_{\mathrm{Dirac, up}}$, whereas for Type~2 charge trapping in this interval can be neglected.  This reversed drift of $V_{\mathrm{Dirac, up}}$ to more negative voltages for slow sweeps results in an increase in the hysteresis width in Type~1 GFETs, see Figure \ref{fig:hyst_width_10}. An increased hysteresis at large sweep ranges for Type~1 confirms our hypothesis, as in Type~1 GFETs $E_\mathrm{F}$ is closer to the Al$_2$O$_3$ defect band. 

In fact, the band alignment shown in Figures~\ref{fig:bands_Type1} and~\ref{fig:bands_Type2} explains the larger hysteresis in Type~1 GFETs in comparison to Type~2 satisfactorily: In Type~1 GFETs biased at $V_\mathrm{Dirac}$, a considerable number of defects is negatively charged. If a negative voltage is applied, these defects discharge due to the band bending and thus $V_\mathrm{Dirac}$ is shifted to more negative voltages during a slow up-sweep (Figure \ref{fig:hyst_Dirac_10}). In Type~2 GFETs, in contrast, the Fermi level is located below the defect band at $V_\mathrm{Dirac}$, as its Fermi level has been shifted down by \SI{200}{meV} via p-doping. Thus, most defects are neutral at the Dirac voltage. If a long time is spent with the GFET biased at negative voltages, the charge states do not change and the location of $V_\mathrm{Dirac}$ during the up sweep is stable independent from the sweep time. 

In short, the higher $\overline{E_\mathrm{T}} - E_\mathrm{F}$ of Type~2 graphene with respect to the Al\textsubscript{2}O\textsubscript{3} defect band leads to a smaller hysteresis width for large sweep ranges. In turn, the threshold voltage in Type~2 GFETs is more stable, as predicted by our stability-based design approach. At small gate bias ranges and fast hysteresis sweeps, Type~2 devices suffer from more charge trapping at the unclean interface with the Al\textsubscript{2}O\textsubscript{3}, and the hysteresis is similar or even higher in Type~2 devices compared to Type~1 (see Figures \ref{fig:hyst_freq_5} and additional data in SI, Figure S4). For fast sweeps, fast traps at the unclean interface in Type~2 GFETs increase the hysteresis, giving the impression of a frequency independent hysteresis width (Figure \ref{fig:hyst_width_10}). Type~1 GFETs exhibit a cleaner interface but a smaller $\overline{E_\mathrm{T}} - E_\mathrm{F}$ with respect to the Al\textsubscript{2}O\textsubscript{3} defects, strongly degrading the GFETs during slow sweeps. For high gate bias ranges and slow sweeps, the border traps of the Al\textsubscript{2}O\textsubscript{3} dominate device stability, thus more stable operation of Type~2 GFETs is observed. These results confirm that it is the alignment of the Al\textsubscript{2}O\textsubscript{3} defect band to graphene $E_\mathrm{W}$ which determines the GFETs' stability and that this alignment can be deliberately tuned by doping the graphene layer.

%
\section{Stability under static gate bias}
\begin{figure}[!htb]
\centering
\vspace*{-0.0cm}
\begin{subfigure}[c]{0.32\textwidth}
\caption{}
\includegraphics[width=\textwidth]{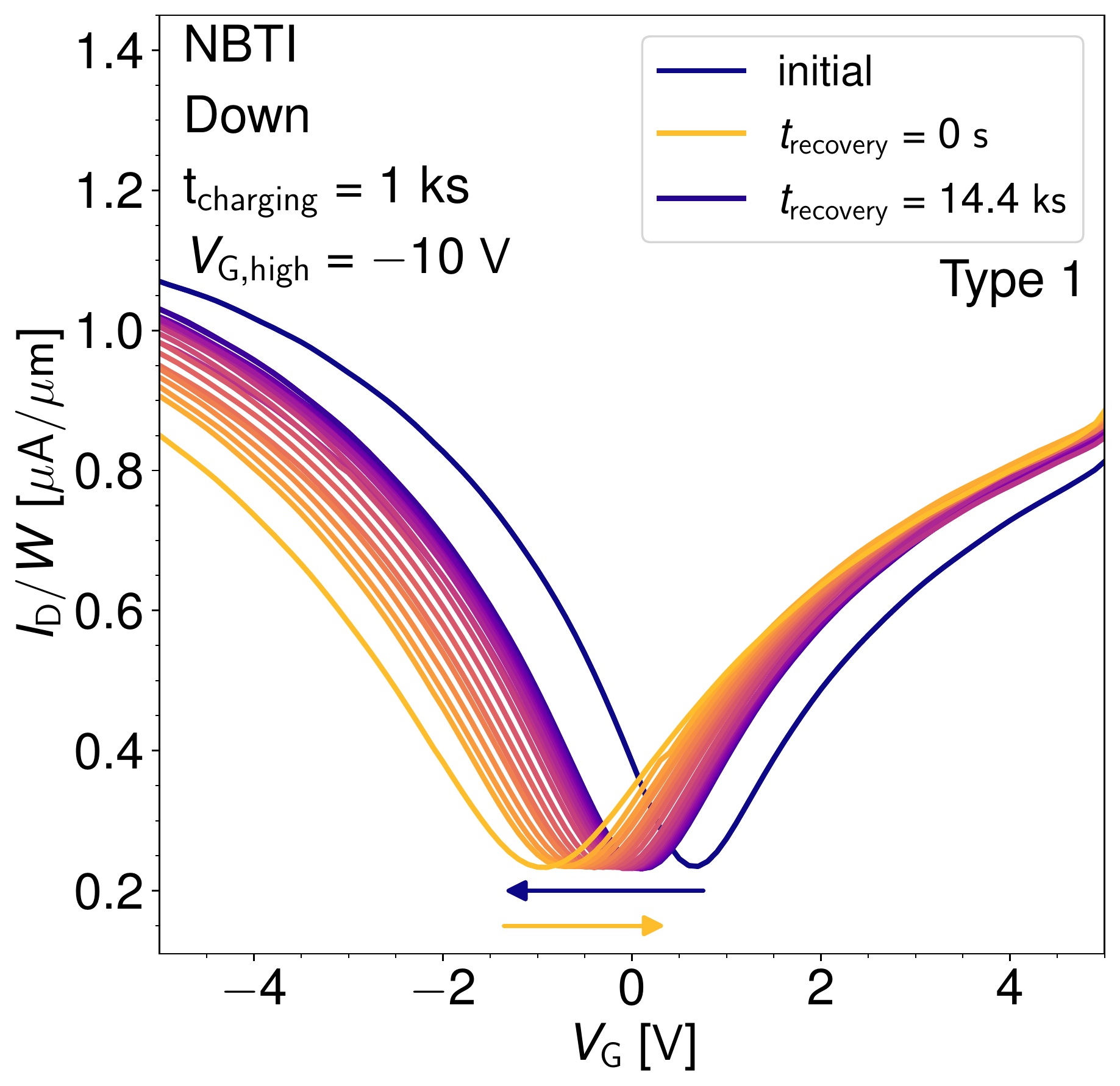}
\label{fig:NBTI_IV}
\end{subfigure}
\begin{subfigure}[c]{0.33\textwidth}
\caption{}
\includegraphics[width=\textwidth]{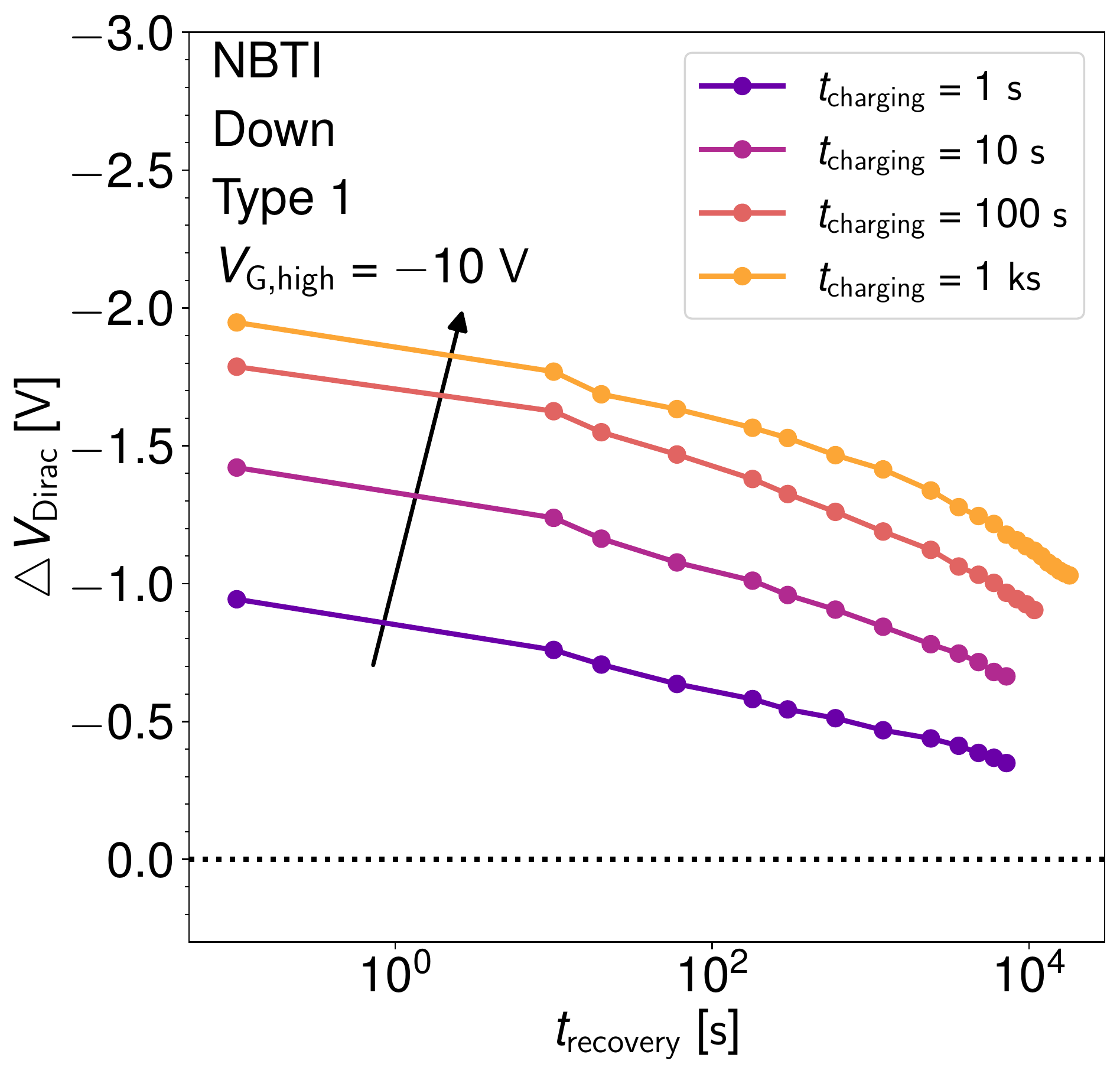}
\label{fig:NBTI_type_1_trace}
\end{subfigure}
\begin{subfigure}[c]{0.33\textwidth}
\caption{}
\includegraphics[width=\textwidth]{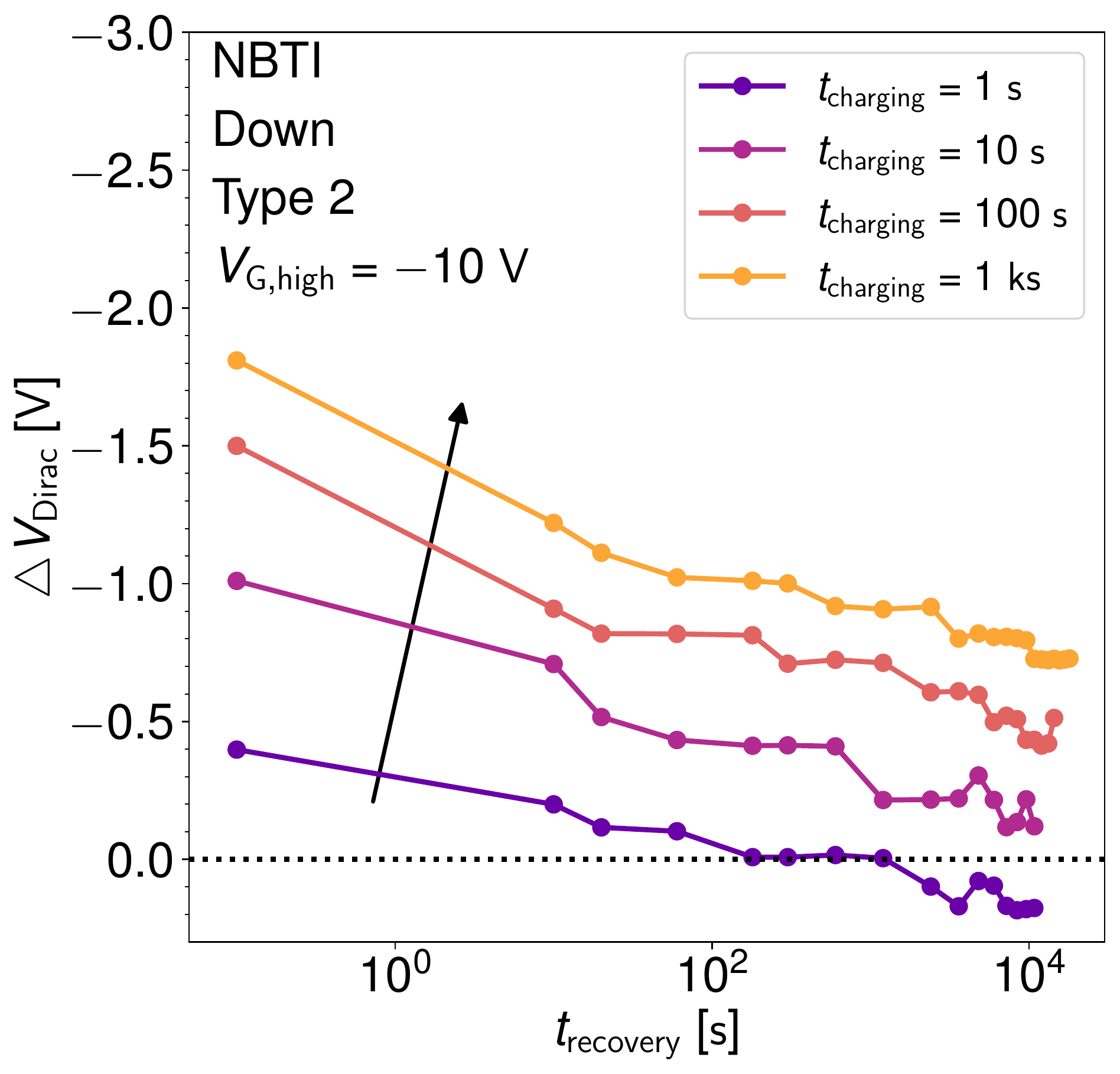}
\label{fig:NBTI_type_2_trace}
\end{subfigure} \\
\begin{subfigure}[c]{0.33\textwidth}
\vspace*{-0.6cm}
\caption{}
\includegraphics[width=\textwidth]{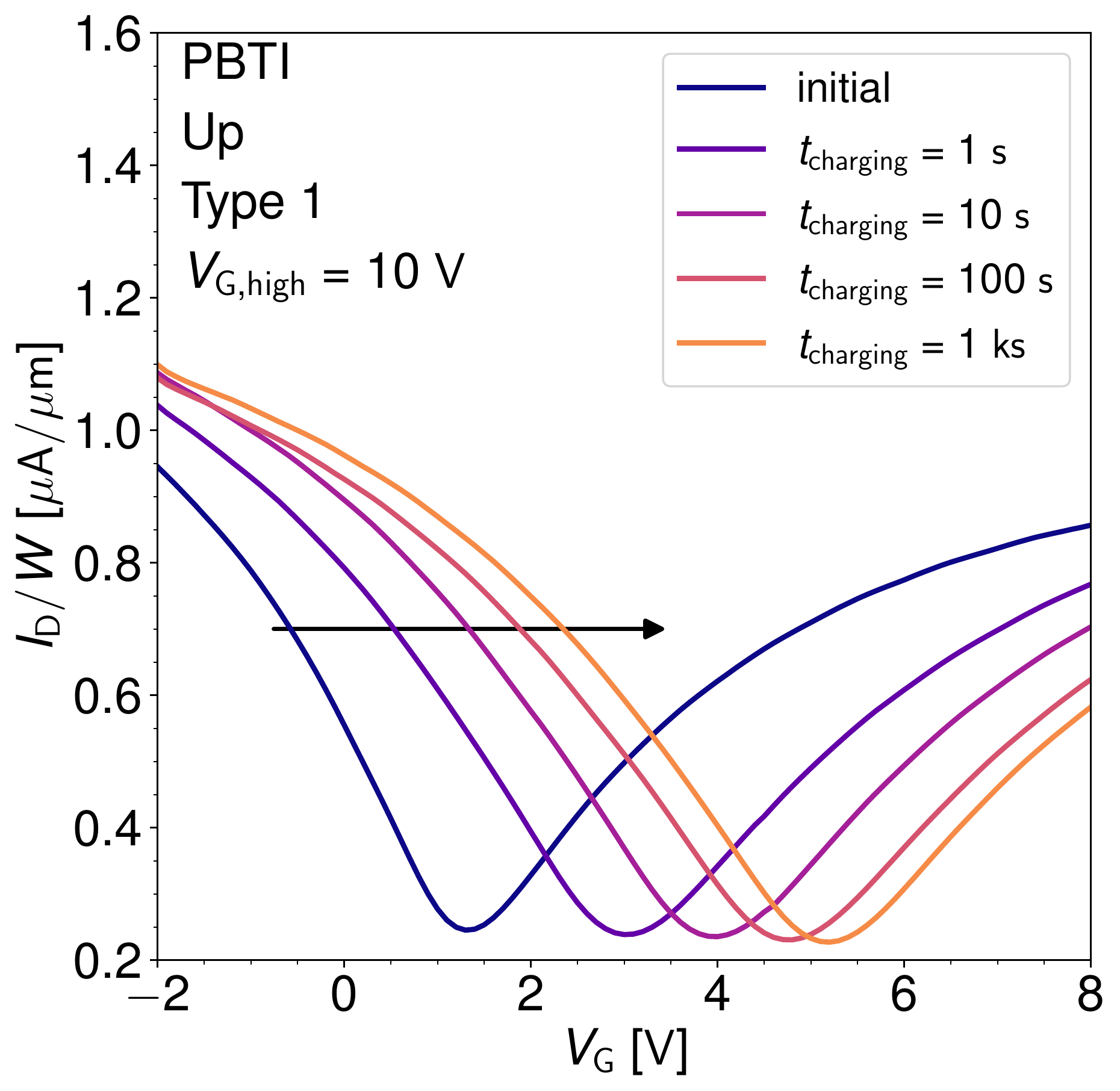}
\label{fig:PBTI_type1_stress_10}
\end{subfigure}
\begin{subfigure}[c]{0.32\textwidth}
\vspace*{-0.6cm}
\caption{}
\includegraphics[width=\textwidth]{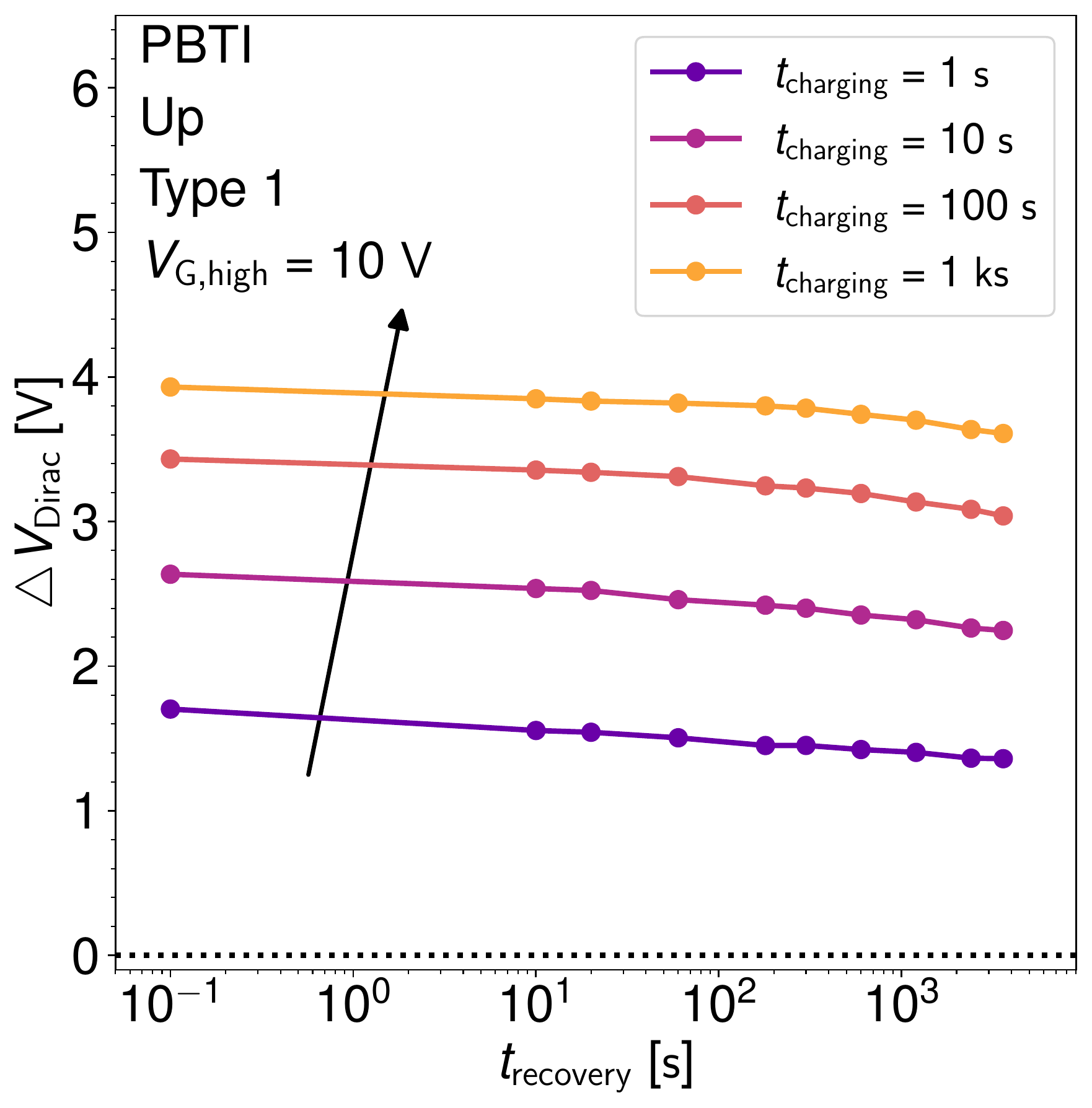}
\label{fig:PBTI_type1_trace}
\end{subfigure}
\begin{subfigure}[c]{0.32\textwidth}
\vspace*{-0.6cm}
\caption{}
\includegraphics[width=\textwidth]{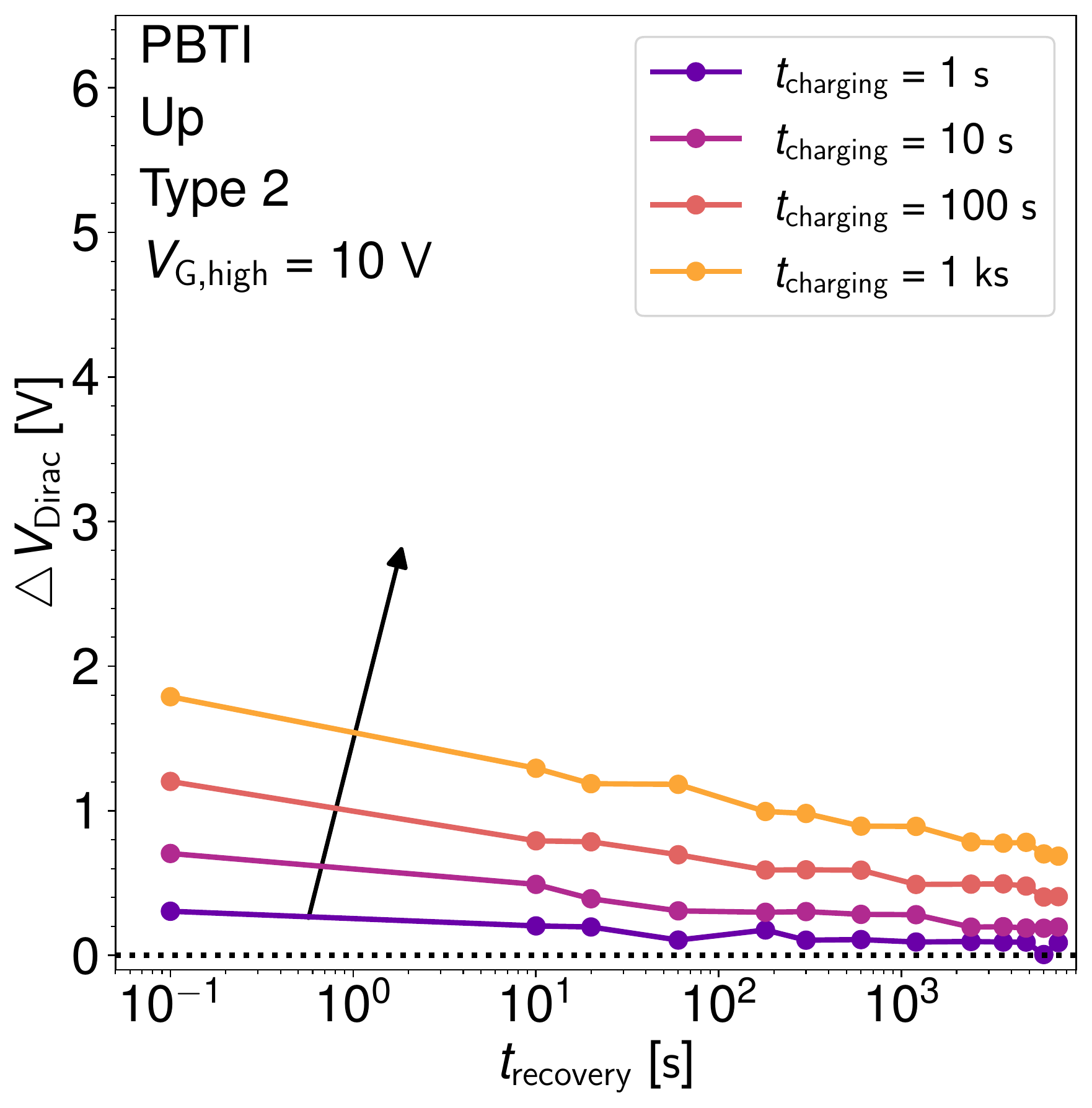}
\label{fig:PBTI_type2_trace}
\end{subfigure}
 \vspace*{-0.6cm}
\caption{\footnotesize In a BTI measurement the FET is subjected to extended periods of elevated gate bias and the drifts of the Dirac voltage during the degradation and recovery periods are recorded, see (a) for a Type~1 device subjected to \SI{-10}{V} for \SI{1}{ks}. 
In (b) a Type~1 FET is subjected for increasing time spans to elevated NBTI gate bias of \SI{-10}{V}, resulting in a larger degradation than observed for the same conditions on Type~2 FETs in (c). When applying to the GFETs an elevated PBTI voltage level of \SI{10}{V} in (d), the Dirac voltage of Type 1 devices drifts more and hardly recovers  (see (e)) in comparison with their Type~2 counterparts (f). In order to avoid an impact of the measurement history, the measurements shown were performed on different devices.}
\end{figure}

For evaluating the long-term stabilty of the GFETs, we analyzed the Dirac voltage shifts ($\Delta V_\mathrm{Dirac}$) after  static elevated gate voltages ($V_{\mathrm{G,high}}$) were applied for varying charging times ($t_{\mathrm{charging}}$). 
We record the magnitude of the initial $\Delta V_\mathrm{Dirac}$ shift and monitor the recovery after the increased gate biasing period with fast $I_{\mathrm{D}}$-$V_{\mathrm{G}}$ sweeps at logarithmically spaced recovery times. In Figure \ref{fig:NBTI_IV} the fast $I_{\mathrm{D}}$-$V_{\mathrm{G}}$ sweeps recorded during the recovery from negative gate biasing (NBTI) at $\SI{-10}{V}$ are shown.

NBTI measured by subjecting the devices to a gate bias of \SI{-10}{V} for increasingly long charging times is shown in Figure~\ref{fig:NBTI_type_1_trace} for Type~1 GFETs and in Figure~\ref{fig:NBTI_type_2_trace} for Type~2 GFETs. As designed, the $V_\mathrm{Dirac}$ shifts are smaller on Type~2 devices than on Type~1 devices. GFETs based on Type~2 graphene are more stable with respect to long-term degradation because graphene's $E_\mathrm{F}$ is further away form the Al\textsubscript{2}O\textsubscript{3} defect band, see Figure~\ref{fig:bands_Type2}. Therefore, on Type~2 GFETs fewer oxide traps change their charge state during negative gate bias, resulting in smaller shifts of $V_\mathrm{Dirac}$ which also recover faster as the traps which emit electrons are located closer to the interface and thus have smaller time constants. For more details, see the recovery traces of NBTI at \SI{-5}{V} in the SI, Figure S5. In Figure~\ref{fig:PBTI_type1_stress_10} the fast $I_{\mathrm{D}}$-$V_{\mathrm{G}}$ sweeps measured after a positive bias at \SI{10}{V} are shown, together with the corresponding recovery traces for Type~1 GFETs in Figure~\ref{fig:PBTI_type1_trace} and for Type~2 GFETs in Figure \ref{fig:PBTI_type2_trace}. For both device types degradation when applying positive biases (PBTI) are higher than NBTI shifts, as the Fermi level in graphene is at the lower edge of the Al\textsubscript{2}O\textsubscript{3} defect band, see Figure \ref{fig:bands_Type1}. Thus, the number of defects which can become more negatively charged during positive bias is larger than the number of defects which can emit one of their electrons during negative bias. As the picture of charge transfer to oxide defects in Al\textsubscript{2}O\textsubscript{3} explains these observations to full satisfaction, these results confirm our stability-based design approach, as we successfully designed Type~2 GFETs to be more stable.By p-doping Type~2 graphene, $E_\mathrm{F}$ was moved further away from the Al\textsubscript{2}O\textsubscript{3} defect band, thus reducing the amount of charge trapping.

Interestingly, throughout all charging times the shifts on Type~1 devices do not recover whereas the shifts on Type~2 devices recover completely. The most surprising observation is that this is also true for a short time of only \SI{1}{s}. This observation was confirmed when subjecting the devices to a smaller gate bias voltage of \SI{5}{V}, see the SI Figure S5. 
To explain the permanent component of BTI degradation, the creation of defects in Al\textsubscript{2}O\textsubscript{3} has to be hypothesized. In silicon FETs using SiO\textsubscript{2} as a gate dielectric the permanent component of BTI has been associated with gate-sided hydrogen release\cite{Grasser2016}. In this model hydrogen diffuses under prolonged biasing conditions through the oxide and creates new oxide defects which cause permanent voltage shifts. We speculate that a similar mechanism of bias facilitated oxide defect creation in the Al\textsubscript{2}O\textsubscript{3} is responsible for the permanent PBTI observed on our GFETs, which will need to be investigated by future studies.

\section{Conclusions}
We have presented the idea that electrically stable FETs based on 2D materials can be designed by tuning the energetic alignment of the Fermi level to reduce the impact of the defect bands in amorphous gate oxides. This approach is founded on the premise that charge trapping at the border traps in amorphous oxides is the key reason which leads to the strong variations in the threshold voltage in 2D FETs and their reduced long-term stability. Based on these facts we suggested a design approach to improve device stability by tuning the Fermi level in 2D materials. In 2D semiconductors the design options mainly lie in choosing suitable materials depending on n- or p-doping or varying the thickness of the layers to minimize the role of known defect bands in the oxides. In graphene, there is more design freedom as the graphene Fermi level can be tuned continuously over a range of up to \SI{2}{eV}. Thus, we demonstrated the validity of our design approach using GFETs with Al\textsubscript{2}O\textsubscript{3} as a top gate oxide and two different types of graphene which only differ in their respective doping and thus their Fermi level alignment. Our measurement results on these two GFET types have shown, that the GFETs based on more p-doped Type~2 graphene with the higher $E_\mathrm{F}$ have a smaller hysteresis and an increased stability of the Dirac voltage when subjected to prolonged elevated gate biases. These results confirm the validity of our stability based design approach and suggest that more stable 2D material based FETs could be built by minimizing the impact of defect bands in the gate oxides in the design process. 

Our approach holds the promise of fabricating electrically stable 2D FETs and is universally applicable to all insulators. We expect that it will lead to further improvements in the electrical stability of devices based on crystalline insulators where the impact of narrow insulator defect bands can be further reduced than in amorphous oxides~\cite{Illarionov2020}. Nevertheless, it remains to be clarified in future studies which levels of electrical stability can be attained with these Fermi level tuned systems based on amorphous oxides and on crystalline insulators, respectively. In addition, stability-based design relies on the knowledge about the defect bands in the oxide which is at the moment incomplete. Thus, it cannot be excluded that in parts of the oxide band gap which is at the moment thought to be free of defect bands, new defect bands might be discovered. Therefore, while the potential gains of taking the stability-based perspective into account from the beginning of the design process could be high, there are currently many unknowns related to feasibility and practicability of the suggested design paradigm which will need to be addressed in future studies.

\section{Methods}

\textit{Device fabrication:} Our top-gated GFETs were fabricated on spin coated polyimide (PI) substrates using photolithography. First, the flexible substrate was prepared by spin coating PI in liquid form on a Si wafer and subsequently curing the layer. The thickness of the solidified PI film was about \SI{8}{\micro m}. During the fabrication process, a rigid Si substrate was used as a support layer. In the next step a CVD grown graphene layer was transferred to the PI substrate. We study two batches of GFETs where the channel is formed by graphene samples purchased from different vendors, namely vendor 1~(Type~1) and vendor 2~(Type~2). For Type~1 devices the CVD graphene was transferred from the copper growth substrate using a PMMA assisted wet transfer method\cite{Suk2011a}, for Type~2 GFETs the transfer was performed by vendor 1. The Type~1 graphene flake covered an area of $2\times\SI{2}{cm}$ and was of higher quality than the Type~1 flake which covered a \SI{6}{inch} wafer. The different quality of the graphene layers was confirmed by Raman spectroscopy, for details see the supporting information. The graphene layer was patterned in an oxygen plasma etch step to form channels of a length ($L$) of \SI{160}{\micro m} and a width ($W$) of \SI{100}{\micro m}. In the next step, the source and drain contacts were deposited by sputtering \SI{50}{nm} Ni, followed by a lift-off process. This step was followed by growing \SI{40}{nm} of Al\textsubscript{2}O\textsubscript{3} with atomic layer deposition~(ALD) on top of the devices to form the gate oxide in a top-gated configuration. In order to finalize the GFETs, the top-gate electrode was fabricated by sputtering \SI{10}{nm} of Ti and \SI{150}{nm} of Al and patterned in a lift-off process. To be able to contact source and drain pads, vias were opened through the Al\textsubscript{2}O\textsubscript{3} with wet buffered oxide etchant.

\textit{Measurement technique:} 
Our electrical measurements were performed in vacuum at room temperature and in complete darkness. The devices were examined with the PI supported on a silicon wafer. From two-probe measurements we extracted the field-effect mobility of the GFETs and found it to be $\SI{4000}{cm^2/Vs}$ for Type~1 graphene and $\SI{1000}{cm^2/Vs}$ for Type~2 graphene. The Hall mobility of both samples was found to be slightly higher. The hysteresis was analyzed by measuring the double sweep $I_{\mathrm{D}}$-$V_{\mathrm{G}}$ characteristics using different sweep times $t_{\mathrm{sw}}$ and sweep ranges $V_{\mathrm{Gminthe}}$ to $V_{\mathrm{Gmax}}$. The hysteresis width $\Delta V_{\mathrm{H}}$ was extracted as the difference between the forward and reverse sweep $V_{\mathrm{Dirac}}$. As was suggested in our previous work~\cite{Illarionov2016f}, we expressed the hysteresis dynamics using the $\Delta V_{\mathrm{H}}$(1/$t_{\mathrm{sw}}$) traces. Finally, the BTI degradation/recovery dynamics were analyzed using subsequent degradation/recovery rounds with either fixed stress time $t_{\mathrm{deg}}$ and increasing high voltage levels $V_{\mathrm{G,high}}$, or fixed $V_{\mathrm{G,high}}$ and increasing $t_{\mathrm{deg}}$. During the recovery period we apply a constant recovery voltage of $V_{\mathrm{G,recovery}} = \SI{1}{V}$ between the sweeps. This voltage is chosen to be close to the charge carrier equilibrium at $V_\mathrm{Dirac}$.  In order to avoid artifacts from fast traps charged during the sweep, the down sweep $I_{\mathrm{D}}$-$V_{\mathrm{G}}$ is used to monitor the recovery of NBTI~\cite{ILLARIONOV14B}. The characteristics obtained when using up sweeps to measure NBTI recovery are shown in the SI Figure S6. For positive bias temperature instability (PBTI) measurements, the recording of the up sweep minimizes artifacts~\cite{ILLARIONOV14B}, thus we used $I_{\mathrm{D}}$-$V_{\mathrm{G}}$ sweeps from negative to positive voltages for the evaluation of PBTI.
As was suggested in our previous study on GFETs~\cite{ILLARIONOV14B}, we expressed the BTI degradation magnitude using Dirac point voltage shift $\Delta V_{\mathrm{Dirac}}$ and plotted it versus the relaxation time $t_{\mathrm{r}}$. In order to gain more statistics, all our measurements were repeated on several devices. 
\vspace*{-0.2cm}



\vspace*{-0.5cm}
\begin{acknowledgement}
\vspace*{-0.5cm}
T.K., Y.Y.I. and T.G. thank for the financial support through FWF grants n$^\circ$ I2606-N30 and n$^\circ$ I4123-N30. T.K. and L.F. gratefully acknowledge financial support through FFG grant n$^\circ$ 1755510.  B.U., Z.W., M.O., D.N. and M.L. acknowledge financial support through DFG grants ULTIMOS2(LE 2440/7-1), GLECSII(NE1633/3-2) and EU grants Graphene Flagship Core3 (881603), WiPLASH (863337). Y.Y.I. and M.W. thank for the financial support through FFG grant n$^\circ$ 867414. Y.Y.I. acknowledges financial support by the Ministry of Science and Higher Education of the Russian Federation under project number 075-15-2020-790. M.W. gratefully acknowledges financial support by the Austrian Federal Ministry for Digital and Economic Affairs and the National Foundation for Research,
Technology and Development.

\end{acknowledgement}
\bibliography{./bib/refs_graphene}



\end{document}